\documentclass[preprint,12pt]{elsarticle}

\usepackage{graphics}
\usepackage{epsfig}

\usepackage{amssymb,amsmath}
\usepackage{color}

\def\be{\begin{equation}}
\def\ee{\end{equation}}
\def\beq{\begin{eqnarray}}
\def\eeq{\end{eqnarray}}

\journal{Comptes Rendus de l'Academie des Sciences}

\begin{document}

\begin{frontmatter}

%% Title, authors and addresses

%% use the tnoteref command within \title for footnotes;
%% use the tnotetext command for the associated footnote;
%% use the fnref command within \author or \address for footnotes;
%% use the fntext command for the associated footnote;
%% use the corref command within \author for corresponding author footnotes;
%% use the cortext command for the associated footnote;
%% use the ead command for the email address,
%% and the form \ead[url] for the home page:
%%
%% \title{Title\tnoteref{label1}}
%% \tnotetext[label1]{}
%% \author{Name\corref{cor1}\fnref{label2}}
%% \ead{email address}
%% \ead[url]{home page}
%% \fntext[label2]{}
%% \cortext[cor1]{}
%% \address{Address\fnref{label3}}
%% \fntext[label3]{}

\title{Numerical simulations of black-hole binaries and gravitational wave emission}

%% use optional labels to link authors explicitly to addresses:
%% \author[label1,label2]{<author name>}
%% \address[label1]{<address>}
%% \address[label2]{<address>}

\author[csic,caltech,centra,olemiss]{Ulrich Sperhake}

\author[caltech,olemiss]{Emanuele Berti}

\author[centra,olemiss]{Vitor Cardoso}

\address[csic]{Institute of Space Sciences (CSIC-IEEC), Campus UAB,
               Torre C5 parells, 08193 Bellaterra, SPAIN}
\address[caltech]{California Institute of Technology, 1200 E California
               Boulevard, Pasadena, CA 91125, USA}
\address[centra]{CENTRA, Departamento de F\'{\i}sica, 
 Instituto Superior T\'ecnico, Universidade T\'ecnica de Lisboa - UTL,
 Av.~Rovisco Pais 1, 1049 Lisboa, Portugal.}               
\address[olemiss]{Department of Physics and Astronomy,
               The University of Mississippi, University, MS 38677, USA}

\begin{abstract}
We review recent progress in numerical relativity simulations of
black-hole (BH) spacetimes. Following a brief summary of the methods employed
in the modeling, we summarize the key results in three major areas
of BH physics: (i) BHs as sources of gravitational waves (GWs),
(ii) astrophysical systems involving BHs, and (iii) BHs
in high-energy physics. We conclude with a list of the most urgent tasks
for numerical relativity in these three areas.
%% Text of abstract

\end{abstract}

\begin{keyword}
%% keywords here, in the form: keyword \sep keyword
Numerical Relativity \sep Black Holes \sep Gravitational Wave Physics \sep
Astrophysics \sep High-Energy Physics

%% MSC codes here, in the form: \MSC code \sep code
%% or \MSC[2008] code \sep code (2000 is the default)
\MSC 65M06 \sep 65M50 \sep 65M70 \sep 65Z05
% 65M06: Finite difference methods
% 65M50: Mesh generation and refinement
% 65M70: Spectral, collocation and related methods
% 65Z05: Applications to physics

\end{keyword}

\end{frontmatter}

%%
%% Start line numbering here if you want
%%
% \linenumbers

%% main text
%=============================================================================
\section{Introduction}
\label{sec:intro}
%For more than half a century after its discovery, the Schwarzschild solution
%for a static, spherically symmetric vacuum spacetime in Einstein's theory
%of General Relativity has been a valuable tool for a wealth of theoretical
%studies, but there remained doubts about the physical relaity of the
%{\em black holes} underlying his solution. 
For almost a century now after its discovery, the Schwarzschild solution
describing a static, spherically symmetric vacuum spacetime in Einstein's
theory of general relativity (GR) has been a valuable tool for a wealth of
theoretical and experimental studies, including in particular weak-field tests
of the theory.  For more than half a century, the validity of the solution
beyond the Schwarzschild radius $r=2GM/c^2$ and the consequent existence of
BHs was thought to be a mathematical artifact of the field
equations.
%Chandrasekhar's upper limit for the mass of a neutron star, beyond which
%collapse ensues, as the star is not able to support the huge gravitational
%forces, was met with skepticism and not taken seriously.
This picture has changed dramatically in more recent decades. Crucial events
in this sense were the mathematical discovery of the Kerr solution describing
rotating BHs \cite{Kerr1963}, and the astrophysical X-ray and radio
observations in the 1960s and 1970s. Cosmic X-ray sources, first detected in
1963 \cite{Giacconi1963}, were soon associated with compact stellar objects
and, in particular, with mass exchange from an ordinary star onto a compact
object in binary systems \cite{Shklovsky1967}. This interpretation received
further support when the X-ray source Cyg X-1 was identified with the
supergiant BOIb star HD~226868 \cite{Bolton1972, Webster1972}, which
established Cyg X-1 as the first observed X-ray binary.  Over the next few
years,
%a number of X-ray binaries
%% with regular eclipses
%were identified,
%% Some of these systems are pulsating sources with delays in
%%synchronization with the eclipses,
%and
the explanation of these systems in terms of accretion onto compact stellar
objects in binary systems became widely accepted (see \cite{VanParadijs1998}
for a review).  Measurements of the binary period and radial velocity
%, combined with estimates for the orbital
%inclination angle from eclipse duration or variable light polarization,
provide lower limits on the mass of the compact companion. The values obtained
for some X-ray binaries comfortably exceed the upper mass limit of about
$3~M_{\odot}$ for neutron stars. Recent investigations of Cyg X-1, for
example, predict $8.7\pm 0.8~M_{\odot}$ for the compact member
\cite{Shaposhnikov2007}.
% Estimates for
%these {\em stellar mass} BHs generally predict masses of the order of
%$10~M_{\odot}$.

Also starting in the early 1960s, optical and radio observations identified
``star like'' objects with surprisingly large redshifts \cite{Matthews1963,
  Schmidt1963}.  If interpreted as cosmological in origin, the redshifts
implied enormous distances and, accordingly, luminosities.  Over the next
decade the cosmological origin of these ``quasi-stellar'' sources (or quasars)
became clear, and accretion onto supermassive BHs (SMBHs) \cite{Salpeter1964,
Zeldovich1964} was accepted as the most plausible explanation of their
energetics \cite{Rees1978,Melia:2007vt}. By now, observations of stellar
dynamics near the centers of galaxies and iron K$\alpha$ emission line
profiles provide strong evidence that many
%if not all
galaxies harbor BHs at their centers \cite{Kormendy1995, Richstone1998}. BH
masses are generally closely correlated to the bulge velocity dispersion and
luminosity \cite{Ferrarese2000, Gebhardt2000}.  For the Milky Way, astrometric
and radial velocity observations of
%the
short-period stars
%S0-2
orbiting the galactic center
%with a period of about $15.2$~years
\cite{Schoedel2002, Ghez2008} predict
%an orbital radius of $8.0\pm0.6~{\rm kpc}$ and
an unseen gravitational mass of $4.1\pm 0.6 \times 10^6~M_{\odot}$.

%\vspace{5mm}
%%%%%%%%%%%%%%%%%%%%%%%%%%%%%%%%%%%%%%%%%%%%%%%%%%%%%%%%%%%%%%%%%%%%%%%%%%%%%%
%\noindent{\bf{\em The need for accurate modeling of black-hole dynamics.}}
%%%%%%%%%%%%%%%%%%%%%%%%%%%%%%%%%%%%%%%%%%%%%%%%%%%%%%%%%%%%%%%%%%%%%%%%%%%%%%
%{\bf[Vitor: next sentence is dangerous: after introducing BHs and
%    seemingly making a strong case for them, we hint that is yet no
%    proof...maybe soften? Also it seems that the spirit would apply to gws but
%    not to BHs...]}
Electromagnetic observations provide strong but {\em indirect} evidence for
the existence of astrophysical BHs. A more direct way of probing the BH nature
of astrophysical compact objects is offered by the newly emerging field of
GW astronomy. First-generation ground-based
laser-interferometric detectors (such as LIGO, VIRGO and GEO600) have reached
design sensitivity, and advanced versions of these detectors will be
operational within a few years \cite{LIGOweb, VIRGOweb, GEO600web}.  One
of the most prominent sources of detectable GWs for these detectors is the
inspiral and coalescence of stellar mass and, possibly, intermediate-mass BH (IMBH) binaries. At
lower frequencies, space-based laser interferometric observations
\cite{ESALISAweb} have the potential to complement ground-based efforts with
high signal-to-noise ratio observations of massive BH binaries. Space-based
detectors are guaranteed to observe galactic binaries containing white dwarfs
and/or neutron stars, and they may also detect the inspiral of stellar compact
objects into massive BHs. Due to the weak interaction of GWs with any type of
matter, including the detectors, digging physical signals from the noisy data
stream represents a daunting task and heavily relies on so-called {\em matched
  filtering} techniques \cite{Finn1992}.  Matched filtering is a common choice
to search for signals of known form in noisy data, and works by
cross-correlating the actual ``signal'' (i.e., the detector's output) with a
set of theoretical templates.  Accurate modeling of the GW sources thus plays
a vital role in maximizing the scientific output from these experimental
efforts.

Event rate estimates for IMBHs and stellar-mass BH
binaries are very uncertain, but advanced Earth-based detectors are expected
to observe several events per year (see \citep{Abadie2010} for a
review). Belczynski {\it et al.} noted that recent metallicity measurements
require revisions in the population synthesis models used for event rate
estimates, suggesting that stellar-mass BH-BH binaries may be the first GW sources to be
detected \cite{Belczynski2010}.  The observations of two possible X-ray binary
precursors of BH-BH binaries provide a much needed observational constraint on
compact binaries containing at least one BH, suggesting again that BH-BH
binary mergers may be more common than anticipated
\citep{Bulik2011}. Population synthesis models have large uncertainties,
related e.g. to the poorly known common envelope phase, and some models would
necessarily be ruled out if Advanced LIGO does {\it not} detect BH
binaries. In this sense, GW detectors are {\em guaranteed} to put constraints
on compact binary formation in the near future. Schutz \cite{Schutz2011} also
pointed out that coherent data analysis using three or more detectors may
further increase event rates by factors of a few, making prospects of doing
astrophysics with Earth-based detectors even brighter. Estimates for
event rates of SMBH binaries detectable by space-based interferometers
range from $0.1$ to $1000$s per year, with a most likely estimate of
$\sim 20-30$ \cite{Volonteri2003,Begelman:2006db,Sesana2007,Arun2008a}.

Over the past decade, BH modeling has also attracted a great deal of interest
in the context of high-energy physics.  According to the gauge/gravity
duality, gravitational physics in anti-de Sitter (AdS) backgrounds describes
field theories on the boundary of that spacetime
\cite{Maldacena1997,Witten1998}. According to this duality, BHs in AdS are
dual to an equilibrium field theory, and therefore interacting BHs or BHs
off-equilibrium may represent interesting physics from the dual point of
view. In fact, there is some evidence that some of the physics behind
quark-gluon plasma formation in heavy-ion collisions is captured by this
duality, which on the gravitational side corresponds to shock wave collisions
and BH formation in AdS \cite{Mateos:2007ay}.

Understanding BH spacetimes is also of interest for high-energy physics in
connection with extra-dimensional scenarios. Solutions to the hierarchy
problem in physics have been developed where all standard-model interactions
are confined to a 3+1 dimensional brane embedded in a spacetime with large
extra dimensions or an extra dimension with a warp factor
\cite{Arkani-Hamed1998, Randall1999}. In these models gravity propagates in
the entire higher-dimensional spacetime (the ``bulk''), and the fundamental
Planck scale could be as low as $1~{\rm TeV}$. These energy scales are
accessible to experiments such as the Large Hadron Collider (LHC).  If the
fundamental Planck scale is so low, Thorne's hoop conjecture suggests the
exciting possibility of BH production in high-energy parton-parton collisions
\cite{Giddings2002,Dimopoulos2001}, i.e.~a direct observable signature of the
existence of extra-dimensions.  Accurate modeling of energy and momentum
losses in the form of GWs during the collisions and determination of the BH
formation cross section will be crucial for the analysis of data from such
experiments (see e.g.~\cite{Frost2009}), and they should be amenable to
classical calculations in $D$-dimensional GR \cite{Banks1999}.

Finally, the understanding of BH dynamics under extreme conditions is
fascinating from a mathematical-physics point of view.  Questions such as BH
stability, cosmic censorship, etcetera can only be addressed by solving the full
nonlinear set of the field equations.

For many of these scenarios, it is convenient to divide the coalescence of a BH
binary in the framework of GR into three stages which mirror the different
tools used for their modeling: (i) The inspiral phase, where the interaction
between the individual holes is still sufficiently weak to justify the use of
post-Newtonian (PN) techniques \cite{Blanchet2006}; (ii) The final orbits,
plunge and merger, which are governed by the strong-field regime of Einstein's
equations and can be described only by fully numerical simulations; (iii) The
ringdown of the remnant BH, which is amenable to a perturbative treatment
\cite{Teukolsky1973, Leaver1986, Berti2009}.

In this article we will focus on the second, strong-field stage, on the
numerical tools dedicated to its study, and results obtained following the 2005 breakthroughs \cite{Pretorius2005a,Campanelli2006,Baker2006}. As we shall see, however, a
comprehensive understanding of BH dynamics requires a close interplay of
numerical and analytical studies, and we will discuss in various cases the
interface between numerical relativity and approximation techniques.

Following a brief summary of the framework employed in numerical relativity
(Sec.~\ref{sec:numerics}), in Secs.~\ref{sec:GWs}-\ref{sec:highenergy} we will
present the main results obtained from numerical studies of BHs in the context
of GW detection, astrophysics and high-energy physics, respectively. We
conclude in Sec.~\ref{sec:conclusions} with a discussion of future
applications of numerical relativity.  At the end of each section we will
provide references to the literature for further reading.

\vspace{3pt}
\noindent
{\bf \em Notation:} We shall be using geometrical units, such that the
gravitational constant $G=1$ and the speed of light $c=1$. We denote
spacetime indices $0\ldots D-1$ by Greek letters, and spatial indices $1\ldots
D-1$ by Latin letters.

%=============================================================================
\section{Numerical modeling of black holes}
\label{sec:numerics}

In Einstein's GR, spacetime is modelled as a $D$-dimensional manifold with a
metric $g_{\alpha \beta}$ of Lorentzian signature $-+++\ldots$. With the exception
of Sec.~\ref{sec:highenergy} we will restrict our discussion to $D=4$,
i.e. three spatial and one time dimension. The metric is determined by the
Einstein equations
\begin{equation}
  G_{\alpha \beta} \equiv R_{\alpha \beta}
     - \frac{1}{2} g_{\alpha \beta}R = T_{\alpha \beta},
     \label{eq:einstein}
\end{equation}
with $T_{\alpha \beta}=0$ in vacuum.
%\vc{Too detailed, I suggest to simply remove: where the Ricci tensor
%$R_{\alpha \beta}$ is a function of the metric and its first and second
%derivatives, $R=g^{\mu \nu}R_{\mu \nu}$ is the Ricci scalar and $T_{\alpha
%\beta}$ the energy-momentum tensor, which vanishes identically in the
%vacuum cases we will be considering here.}
While analytic solutions (as, for example, the Schwarzschild and Kerr
solutions and the Friedmann-Lema{\^i}tre-Robertson-Walker solution) have been
found for systems with high symmetry, solutions for dynamical configurations
without special symmetries generally require the use of numerical tools.  For
this purpose the Einstein equations need to be cast as an {\em initial value
  formulation} such that the Cauchy-Kowalewski theorem guarantees a uniquely
determined evolution of appropriately chosen initial data. In the case of
Einstein's equations, it is not obvious that such a formulation can be
obtained: a straightforward calculation shows that Eq.~(\ref{eq:einstein})
contains second time derivatives of the metric components $g_{ab}$, but not of
$g_{t\alpha}$.  This fact mirrors the coordinate or gauge freedom
characteristic of GR, and implies the existence of so-called {\em constraint}
equations which impose conditions on the metric components and their first
derivatives on each spatial slice with constant coordinate time $t$. In the
following we will discuss two approaches which provide an initial value
formulation of the Einstein equations and have led to successful, long-term
stable numerical evolutions: the generalized harmonic (GH) formulation
employed in Pretorius' breakthrough BH binary simulations
\cite{Pretorius2005a} and the canonical Arnowitt-Deser-Misner (ADM) split
\cite{Arnowitt1962}, reformulated by York \cite{York1979}, which forms the
starting point for the {\em moving puncture} breakthroughs
\cite{Campanelli2006, Baker2006}.

\vspace{3pt}
\noindent
{\bf \em ADM 3+1 Split:} Spacetime is decomposed into a one-parameter family of
three-dimensional spatial slices. We choose coordinates adapted to this
formulation, such that the coordinate time $t$ labels each slice and $x^i$
denotes points on the slice.
%
%\eb{I added ``denote'' - is this what you meant?}
%
On each hypersurface, there exists a unique timelike, future pointing unit
normal field $n^{\alpha}$ which defines a projection operator
$\bot^{\alpha}{}_{\beta}=g^{\alpha}{}_{\beta}+n^{\alpha} n_{\beta}$ onto the
hypersurface. The geometry of the hypersurface is completely determined by the
{\em three-metric} or {\em first fundamental form} $\gamma_{ij}\equiv
\bot^{\mu}{}_i \bot^{\nu}{}_j g_{\mu \nu} = g_{ij}$.  The coordinate freedom
is represented by the shift vector $\beta_i \equiv g_{ti}$ and the lapse
function $\alpha \equiv \sqrt{g_{tt}-\beta^m \beta_m}$ (with $\beta^i \equiv
\gamma^{im} \beta_m$), which relate spatial coordinates on (and measure
separation in proper time between) different hypersurfaces. The projections
$\bot^{\alpha}{}_i \bot^{\beta}{}_j G_{\alpha \beta}$, $\bot^{\alpha}{}_i
G_{\alpha \mu} n^{\mu}$, $G_{\mu \nu } n^{\mu}n^{\nu}$ of the Einstein
equations then lead to six evolution equations for the three-metric
$\gamma_{ij}$, three momentum constraints and the Hamiltonian constraint:
\begin{eqnarray}
  \partial_t \gamma_{ij} &=& \beta^m \partial_m \gamma_{ij}
      + \gamma_{mi} \partial_{j} \beta^m
      + \gamma_{mj} \partial_{i} \beta^m
      - 2\alpha K, \\
  \partial_t K_{ij} &=& \beta^m \partial_m K_{ij}
      + K_{mi} \partial_{j} \beta^m
      + K_{mj} \partial_{i} \beta^m
      - D_i D_j \alpha \nonumber \\
   && + \alpha \left(
        \mathcal{R}_{ij} + KK_{ij} - 2K_{im}K^{m}{}_j
        \right), \\
  \mathcal{H} &\equiv& \mathcal{R} + K^2 - K_{mn}K^{kn} = 0, \\
  \mathcal{M}^i &\equiv& D_m \left(
        K^im - \gamma^{im}K
        \right) = 0,
\end{eqnarray}
where $\mathcal{R}_{ij}$, $\mathcal{R}$ and $D_i$ denote the Ricci tensor,
Ricci scalar and covariant derivative associated with the three-metric, and the
extrinsic curvature $K_{ij}$ has been introduced to obtain a first-order
system in time.

It turns out that this formulation of the Einstein equations is only {\em
weakly hyperbolic} \cite{Alcubierre2008},
%
%\eb{What is a ``weakly'' hyperbolic system?}
%
and therefore not suitable for long-term stable numerical
evolutions. Motivated by these stability problems, several modifications of
the ADM system have been investigated.  The most popular is the
Baumgarte-Shapiro-Shibata-Nakamura (BSSN) formulation \cite{Shibata1995,
  Baumgarte1998}, which rearranges the degrees of freedom via a split of the
extrinsic curvature into trace and trace-free parts, a conformal
transformation and promotion of the contracted conformal Christoffel symbols
to the status of independent variables:
\begin{eqnarray}
  && \phi = \frac{1}{12} \ln \gamma, \,\,\,\,\,\,\,\,\,\,\,
       \tilde{\gamma}_{ij} = e^{-4\phi} \gamma_{ij}, \nonumber \\
  && K = \gamma^{ij}K_{ij}, \,\,\,\,\,\,\,\,\,\,
       \tilde{A}_{ij} = e^{-4\phi} \left(
       K_{ij}-\frac{1}{3} \gamma_{ij} K \right), \nonumber \\
  &&\tilde{\Gamma^i} = \tilde{\gamma}^{mn}\tilde{\Gamma}^i_{mn}
      = \frac{1}{2}\tilde{\gamma}^{mn} \tilde{\gamma}^{ik}
        \left(
          \partial_m \tilde{\gamma}_{nk}
        + \partial_n \tilde{\gamma}_{km}
        - \partial_k \tilde{\gamma}_{mn}
        \right).
  \label{eq:BSSN}
\end{eqnarray}
The resulting set of evolution and constraint equations is given in
Eqs.~(15)-(30) of \cite{Alcubierre2003b}.  Subject to minor modifications,
such as replacing the conformal factor by $\chi \equiv e^{-4\phi}$ or adding
the auxiliary constraint arising from the definition of $\tilde{\Gamma}^i$ in
Eq.~(\ref{eq:BSSN}), this formulation is employed in the present generation of
moving puncture evolution codes \cite{Baker2006a,Campanelli2006a,Herrmann2006,
  Sperhake2006,Bruegmann2006a,Pollney2007,Liu2009} which fix the gauge
via ``1+log'' slicing and the $\Gamma$-driver condition \cite{Alcubierre2003b,
vanMeter2006,Mueller2010,Alic2010}.

\vspace{3pt}
\noindent
{\bf \em GH formulation:} Harmonic coordinates are defined by the condition
$g_{\alpha \mu} \Box x^{\mu} = -\Gamma_{\alpha}=0$, where $\Box \equiv
\nabla^{\mu} \nabla_{\mu}$ represents the scalar wave operator. These
coordinates have played an important role in the analysis of the Cauchy
problem in GR \cite{Bruhat1952, Bruhat1962, Fischer1972}. In harmonic
coordinates, the Einstein equations take on a manifestly hyperbolic form,
which allows for a generalization to arbitrary coordinate or gauge choices
\cite{Friedrich1985, Garfinkle2002}. The first step is to introduce four
arbitrary source functions $H^{\alpha}$ such that the coordinates obey
\begin{equation}
  -\Gamma^{\alpha} = \Box x^{\alpha} = H^{\alpha},
     \label{eq:H}
\end{equation}
and treat these functions as independent evolution variables. Then one
considers the GH system
\begin{equation}
  R_{\alpha \beta} - \nabla_{(\alpha} C_{\beta)} = 0,
      \label{eq:GH}
\end{equation}
where $C_{\alpha} \equiv H_{\alpha} + \Gamma_{\alpha}$. Eq.~(\ref{eq:GH}) is
equivalent to the Einstein equations, subject to the validity of the
constraint (\ref{eq:H}). In expanded form, the GH system is given by
\begin{equation}
  g^{\mu \nu} \partial_{\mu} \partial_{\nu} g_{\alpha \beta} =
      - 2\nabla_{(\alpha} H_{\beta)}
      +2 g^{\mu \nu} g^{\kappa \lambda}
      \left(
        \partial_{\kappa} g_{\mu \alpha} \partial_{\lambda}g_{\nu \beta}
        - \Gamma_{\alpha \mu \kappa} \Gamma_{\beta \nu \lambda}
      \right),
    \label{eq:GHexpanded}
\end{equation}
and the constraints $C_{\alpha}=0$ are preserved by virtue of the Bianchi
identities provided that $C_{\alpha}$ and $\partial_t C_{\alpha}$ vanish on
the initial hypersurface $t=0$. A key ingredient in the numerical evolution of
the GH system is the addition of a constraint damping term $\propto \left[
  2\delta^{\mu}{}_{(\alpha} t_{\beta )} -g_{\alpha \beta} t^{\mu} \right]
C_{\mu}$ to the right hand side of Eq.~(\ref{eq:GHexpanded})
\cite{Gundlach2005}. Here, $t^{\mu}$ is a non-vanishing timelike vector field.
Finally, it is interesting to note that the Hamiltonian
and momentum constraints are automatically satisfied on the initial
hypersurface if $C_{\alpha} = 0 = \partial_t C_{\alpha}$ \cite{Lindblom2005}.

The coordinates were determined in Pretorius' initial simulations
by
\begin{equation}
  \Box H_t = -\xi_1 \frac{\alpha-1}{\alpha} + \xi_2 n^{\nu}
       \partial_{\nu} H_t, \,\,\,\,\,\,\,\, H_i = 0,
\end{equation}
and the evolution of the metric proceeds according to
Eq.~(\ref{eq:GHexpanded}) with the constraint damping term of Gundlach {\em et
  al.} \cite{Gundlach2005}. For further discussion of gauge choices in the
GHG system, see also \cite{Lindblom2009a,Choptuik2009}.

The spectral code originally developed by the Caltech-Cornell group (see
\cite{Scheel2006, Boyle2007} and references therein) employs a first-order
version of the GH system \cite{Lindblom2005} with dual-coordinate frames and
BH excision. Furthermore, the first-order GH formulation facilitates the
specification of constraint-preserving boundary conditions
\cite{Lindblom2005}. During the inspiral phase, they evolve the gauge
functions $H_{\alpha}$ such that they remain constant in a frame comoving
with the BHs, but switch to a dynamic evolution during the plunge and
ringdown; see Sec.~III in \cite{Chu2009}.  While more complex in structure,
this framework enables their code to generate BH evolutions with exceptional
accuracy \cite{Boyle2008,Chu2009}.

\vspace{3pt}
\noindent
{\bf \em Initial data:} The construction of initial data requires solving the
Hamiltonian and momentum constraints. Here we only summarize the key
concepts; we refer the reader to Cook's review article \cite{Cook2000} for
details. Most work on the construction of initial data is based on the
York-Lichnerowicz split \cite{Lichnerowicz1944,York1979}, which involves a
conformal transformation of metric and extrinsic curvature and separates the
latter into trace and trace-free part. More recently, the {\em thin-sandwich}
construction \cite{York1999}, which replaces the extrinsic curvature by the
time derivative of the metric, has become a popular alternative. In either
case, the resulting elliptic equations simplify substantially under the
assumption of conformal flatness and a spatially constant $K$.

While this formalism provides a convenient method to solve the constraint
equations, we still need to ensure that the initial data represent a
physically realistic system, typically two BHs with specific spins and
momenta. This can be achieved by generalizing the Schwarzschild solution,
which is obtained in the above framework in conformally flat form with
conformal factor $\exp(\phi)=1+\frac{m}{2r}$. A generalization to initial data
of $N$ BHs starting from rest is directly obtained by the construction of
Misner \cite{Misner1960} or Brill and Lindquist \cite{Brill1963}. Remarkably,
an analytic solution for the extrinsic curvature can still be obtained for BHs
with initial linear momenta $\mathbf{P}_n$ and spins $\mathbf{S}_n$
\cite{Bowen1980}. By applying a compactification to the internal asymptotically
flat region, Brandt \& Br{\"u}gmann \cite{Brandt1997} arrived at the
so-called {\em puncture data} construction,
which is the method of choice for most numerical evolutions employing the
BSSN formulation.  In order to overcome an upper limit of $\approx 0.93$ for
the dimensionless spin of BHs in puncture type initial data \cite{Dain2008},
Lovelace {\em et al.} \cite{Lovelace2010} generated initial data based on a
generalization of the Kerr-Schild solution \cite{Kerr1965} and were thus able
to evolve a BH binary with spin magnitude 0.95.

In alternative to generalizing analytically known BH solutions, the
presence of horizons in the initial data can be encoded in the form of
boundary conditions for the metric and extrinsic curvature \cite{Cook2004,
  Dain2004} as determined by the isolated horizon framework
\cite{Ashtekar2004}.  Initial data along these lines have been constructed in
\cite{Cook2004, Dain2004}, and form the starting point of most of the numerical
evolutions using the GH system (see e.g. \cite{Buonanno2006,Boyle2007}).

\vspace{3pt}
\noindent
{\bf \em Diagnostics:} Extracting physical information from numerical
simulations of the Einstein equations is nontrivial for two reasons.  First,
the evolved variables are dependent on the coordinate conditions; second, it
is often not possible to define local quantities familiar from Newtonian
physics. The first difficulty requires the calculation of gauge-independent
variables. The second difficulty is alleviated by the {\em isolated horizon}
framework \cite{Ashtekar2004}, which facilitates the calculation of BH mass
and spin in the limit of isolated BHs; in practice, this framework is
often applied when the BHs are farther apart and their interaction is
considered negligible.
%
%\eb{Can this be made more precise?}
Local properties of the BHs are encoded in their apparent horizon
\cite{Baumgarte1996,Anninos1998a,Thornburg2004,Krishnan2007}.  In particular,
the irreducible mass can be expressed in terms of the apparent horizon area:
$M_{\rm irr}=\sqrt{A_{\rm AH}/16\pi}$. In the limit of an isolated BH, the
spin can be derived from the integration of the rotational Killing vectors
over the horizon according to Eq.~(8) of \cite{Krishnan2007}. In practice it
has been found that using {\it flat-space} rotational Killing vectors yields
reasonable results \cite{Campanelli2006d}, but we also note the improved
method to compute approximations to Killing vectors in \cite{Cook:2007wr}
and an alternative method \cite{Jasiulek:2009zf}
based on Weyl scalars to compute quasi-local BH quantities
without solving the Killing equation. It is important to bear in mind,
however, the approximate nature of any spin calculation in BH binary
simulations.

In comparison, it is more straightforward to extract global quantities of the
spacetime, most notably the total mass-energy $M_{\rm ADM}$ and the linear and
angular momentum $P_i$, $J_i$ \cite{Arnowitt1962,York1979}.  These are given
by surface integrals at spatial infinity, see e.g.~Eqs.~(7.15), (7.56) and
(7.63) in \cite{Gourgoulhon2007}.
%%
%\begin{eqnarray}
%  M_{\rm ADM} &=& \frac{1}{16\pi} \lim_{r\rightarrow \infty}
%      \int_{S_r} \sqrt{\gamma} \gamma^{ij} \gamma^{kl}
%      (\partial_j \gamma_{ik} - \partial_k \gamma_{ij})
%      dS_l, \\
%  P_i &=& \frac{1}{8\pi}\lim_{r\rightarrow \infty}
%      \int_{S_r} \sqrt{\gamma}(K^m{}_i - K \delta^m{}_i ) dS_m, \\
%  J_i &=& \frac{1}{8\pi}\epsilon_{il}{}^m \lim_{r\rightarrow \infty}
%      \int_{S_r} \sqrt{\gamma}x^l K^n{}_m - K\delta^n{}_m),
%\end{eqnarray}
%%
%where $\epsilon_{ijk}$ is the Levi-Civita symbol and $S_r$ denotes a
%coordinate sphere of radius $r$.

Arguably the most important diagnostic quantity in BH simulations is the GW
signal. The most common method to extract GWs is based on the Newman-Penrose
formalism \cite{Newman1962} and derives the complex Newman-Penrose scalar
$\Psi_4$ from contraction of the Weyl tensor with suitably chosen tetrad
vectors \cite{Kinnersley1969}, Sec.~III A in \cite{Bruegmann2006a}. It is often convenient to decompose $\Psi_4$ into multipoles
$\psi_{lm}$ using spherical harmonics of spin-weight $-2$:
$\Psi_4(t,r,\theta,\phi) = \sum_{l=2}^{\infty} {}_{-2}Y_{lm}(\theta, \phi)
\psi_{lm}(t,r)$.  Contemporary numerical codes typically evaluate $\Psi_4$ at
finite coordinate radius. This results in systematic errors, due to ambiguities
in the tetrad choice and the neglect of GW backscattering. While these errors
can be reduced by extracting GWs at a sequence of radii and extrapolating to
infinite radius ($\Psi_4$ asymptotically falls off $\sim r^{-1}$
\cite{Hinder2011}), a cleaner method is to calculate $\Psi_4$ at infinity as
provided by {\em Cauchy-characteristic extraction} methods
\cite{Reisswig2009,Reisswig2009a,Babiuc2010}. From $\Psi_4$, one
straightforwardly obtains the energy, linear and angular momentum radiated in
the form of GWs; see e.g.~Eqs.~(49)-(51) in \cite{Bruegmann2006a}. In GW data
analysis it is more common to work with the wave strain
\begin{equation}
  h \equiv h_+ - ih_{\times} = \int_{-\infty}^t dt' \int_{-\infty}^{t'}
      dt'' \Psi_4,
\end{equation}
decomposed into multipoles:
%\vc{I don't like the $H_{lm}$ notation and would prefer $h_{lm}$, also
%for consistency with $\Psi_4$ and because it is similar to the GHG
%variable. However, if it has been used before I am OK}
$h(t,r,\theta,\phi)
=\sum_{l=2}^{\infty}{}_{-2}Y_{lm}(\theta,\phi)h_{lm}(t,r)$.  In practice,
after integrating $\Psi_4$ twice in time the signal can be severely affected
by nonlinear drifts, which can be controlled to a significant extent by
performing the integration in the frequency domain \cite{Vaishnav2007,
  Reisswig2010}.

Alternatively, GWs can be extracted by
viewing the metric in the far field regime as a perturbation of the
Schwarzschild spacetime and employing the formalism of Regge, Wheeler
\cite{Regge1957} and Zerilli \cite{Zerilli1970, Zerilli1970b}. One thus
obtains the Regge-Wheeler-Moncrief and Zerilli-Moncrief master functions
$Q^{\times}_{lm}$, $Q^+_{lm}$, which can be converted into the multipolar
components of the GW strain $h$ according to Eq.~(49) of \cite{Reisswig2010a}.

For a comprehensive summary of the numerical framework for evolving Einstein's
field equations, we refer the reader to the books by Alcubierre
\cite{Alcubierre2008} and Baumgarte and Shapiro \cite{Baumgarte2010}.
%For a brief summary of numerical methods, see \cite{McWilliams2010}.
Further details, including mathematical aspects of the equations and
characteristic techniques, can be found in the review articles
\cite{Lehner2001,Gourgoulhon2007,Jaramillo2007,Pfeiffer:2012pc}.

%(i) Generalizations of the Schwarzschild solution. As expected, the
%Schwarzschild solution is obtained in the above mentioned framework
%in the form of isotropic coordinates, i.e.~a conformally flat metric
%and a conformal factor $exp(\phi)=(1+\frac{m}{2r})$.

%=============================================================================
\section{Gravitational wave physics}
\label{sec:GWs}

%The weak interaction of GWs with any type of matter is the double-edged sword
%of GW physics. By propagating essentially undisturbed, GWs provide us with the
%cleanest signature of their sources, which more than justifies the often used
%phrase of {\em opening up a new window to the Universe}. Unfortunately ``all
%types of matter'' includes the GW detectors themselves, and great
%technological advances were necessary to achieve the extreme precision
%measurements required for GW observation.

BH binary systems represent one of the most promising sources of detectable
GWs. The parameters of a BH binary are commonly divided into {\em intrinsic}
and {\em extrinsic} \cite{Buonanno2003}. Intrinsic parameters characterize
physical properties of the system, such as the total mass $M$, the mass ratio
$q\equiv M_2/M_1\le 1$, the individual BH spins ${\bf S}_1$, ${\bf S}_2$ and the
orbital eccentricity.  Extrinsic parameters such as sky position, distance,
orbital inclination angle, arrival time and initial phase of the wave, in
contrast, depend on the source location relative to the observer, and do not
directly enter the GW source modeling process.

The majority of numerical studies of BH binary spacetimes performed to date
have focussed on comparable mass-ratios $q \ge 1/10$ and moderate spin
magnitudes $\chi_i \equiv |{\bf S}_i|/M_i^2 \lesssim 0.8$, with particular
emphasis on spins (anti-)aligned with the orbital angular momentum. We note,
however, the following explorations outside this ``best charted'' subset of
the parameter space. Circular binaries with mass ratios up to $q=1/100$ have
been studied in \cite{Lousto2010c,Lousto2010b}; comparisons for this value of
$q$ with fully perturbative calculations have been obtained in the limit of
head-on collisions in \cite{Sperhake2011}, who report a discrepancy of $\sim
7~\%$ in GW energy and momentum, probably in large part due to the
discretization error of the numerical simulations.
The first direct comparison of numerical results with leading-order
self-force prediction was studied in the context of periastron advance
in BH binaries in \cite{LeTiec2011}.  BH binaries with spins
$\sim 0.9$ have been evolved in \cite{Marronetti2007a,Dain2008,Lovelace2008},
but exceeding an apparent barrier of $\chi\approx 0.93$
\cite{Lousto:2012es} requires departure
from the conformal flatness assumption
\cite{Hannam2007a,Liu2009,Lovelace2010}. The present maximum is
$\chi_{1,2}=0.95$ for aligned spins evolved for 12.5 orbits by Lovelace {\em
  et al.} \cite{Lovelace2010}. Finally, we note that emission of GWs
efficiently circularizes BH binary orbits \cite{Peters:1963ux}. BH binaries
are therefore expected to have vanishing eccentricity by the time they enter
the frequency window of ground-based detectors, and for this reason most work
on GW source modeling has focused on the quasi-circular limit. The derivation
of BH momenta, including a small radial component, for quasi-circular initial
data in numerical relativity is based either on integrating the PN equations
of motion \cite{Husa2007,Walther2009} or iterative procedures using several
numerical simulations \cite{Pfeiffer2007,Boyle2007,Mroue2010,Tichy2011,
Purrer:2012wy}.
The decay of eccentricity during the inspiral phase as well as
periastron advance has been measured
in numerical simulations in \cite{Mroue2010}.
For numerical studies of BH binaries with significant eccentricity as well as
arguments why eccentric binaries may represent relevant sources of GWs after
all, especially in the more extreme mass ratio regime and for space-based
detectors, we refer the reader to
\cite{Sperhake2007,Hinder2008,AmaroSeoane2007,Yunes:2009yz,Sesana:2010qb}.

\begin{figure}[t]
\centering
\includegraphics[height=6.3cm,clip=true]{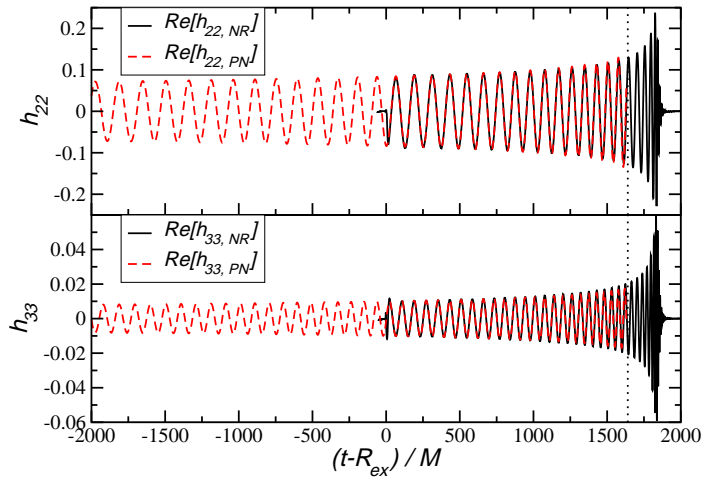}
\caption{Real part of the GW multipoles $h_{22}$ and $h_{33}$ obtained from
  the inspiral and merger of a nonspinning binary with mass-ratio $q=1/4$. The
  dashed red curve represents PN predictions matched to the numerical (solid
  black) signal. The vertical dotted lines mark the time when the frequency of
  the $(l,m)=(2,2)$ multipole has reached $M\omega=0.1$.}
\label{fig:Hlm}
\end{figure}

\vspace{3pt}
\noindent
{\bf \em Gravitational waveforms from BH binaries:} For illustration, we
display in Fig.~\ref{fig:Hlm} the real parts of the $h_{22}$ and $h_{33}$
multipoles of the GW strain $h$ obtained for the inspiral and coalescence of a
binary
of nonspinning BHs with mass ratio $q=1/4$ \cite{Sperhake2010a}.  In the course
of the inspiral, both amplitude and frequency of the GW signal gradually
increase. Close to merger, defined as formation of a common apparent horizon,
the GW amplitude reaches a maximum. Eventually it drops exponentially as
the merged hole rings down to a stationary Kerr state: see e.g.~Fig.~18 in
\cite{Buonanno2006}.

The qualitative features of the GW signal emitted by different types of
binaries can be summarized as follows.  (i) The inspiral of two nonspinning,
equal-mass BHs is the case most intensively studied by numerical
relativity. The wave signal is dominated by the $(l,m)=(2,2)$ multipolar
component, which carries $>98~\%$ of the total radiated energy. The longest
numerical simulation by Scheel {\em et al.} \cite{Scheel2008} covers 16 orbits
and reports a ratio of final to initial BH mass $M_f/M=0.95162\pm 0.00002$ and
a final spin $S_f/M_f^2=0.68646\pm 0.00004$.  (ii) As the mass-ratio
decreases, higher-order modes become more prominent \cite{Berti2007} and the
fractional energy in $(l,m)=(2,2)$ drops to about $68~\%$ in the limit $q
\rightarrow 0$ \cite{Gonzalez2008}. Motivated by the strong relation
of phase and frequency of different $(l,m)$ multipoles,
Baker {\em et al.} \cite{Baker2008a} view
the $(l,m)$ radiation modes as being generated by the corresponding momenta
of an {\em implicit rotating source}; see also \cite{Kelly2011}.
(iii) GW multipoles with odd $l$
%\vc{I think we mean odd $l$, or we should mean. All odd $l$ quantities
%are suppressed in the equal-mass limit}
are suppressed in the equal-mass limit due to symmetry.  (iv) BH binaries with
spins aligned with the orbital angular momentum emit stronger GW signals due
to an effect sometimes referred to as {\em orbital hang-up}: equal-mass
binaries with $\chi_1=\chi_2=0.757$ radiate about twice as much energy and
angular momentum when compared to the nonspinning case
\cite{Campanelli2006b}. For anti-aligned spins the GW energy decreases by a
similar factor.  (v) Orbital eccentricity introduces a nonmonotonic behavior
of the GW frequency; cf.~Fig.~6 in \cite{Hinder2008}.  (vi) Spin-spin and
spin-orbit couplings cause a precession of the orbital plane in the case of BH
binaries with spins that are not aligned or anti-aligned with the orbital
angular momentum. This precession manifests itself in a modulation of the GW
amplitude emitted in a fixed angular direction: cf.~Fig.~1 in
\cite{Kidder1995}.

\vspace{3pt}
\noindent
{\bf \em Tests of general relativity:} 
Numerical simulations of binary BH mergers provide an opportunity to study the
nonlinearities of the theory and to test the Kerr nature of astrophysical
BHs. If GR is the correct theory of gravitation, all BHs
in the Universe should be described by the Kerr solution \cite{Kerr1963},
which depends on two parameters: the mass $M$ and the dimensionless spin
$\chi$. The remnant of a binary BH merger settles down to the Kerr solution by
emitting gravitational radiation at characteristic (complex) quasinormal
frequencies $M\omega_{nlm}$ that depend only on $\chi$. Here $(l,m)$ are the
usual angular indices, and $n$ is the ``overtone number'': modes with small
$n$ have a longer damping time and should dominate the radiation
\cite{Leaver:1985ax,Berti2009}.

The measurement of the real and imaginary part of a single quasinormal mode
contains, in principle, enough information to determine the mass and spin of
the BH. Together with the accurate determination of the individual BH masses from the inspiral phase 
this already allows for tests of GR, by testing the GR prediction for mass loss during inspiral and merger and at the same time the strong field dynamics of GR. The measurement of {\it any} other frequency or damping time
then provides a further interesting test of the Kerr nature of the final BH \cite{Dreyer:2003bv}.  The
feasibility of these tests depends on the characteristics of a given GW
detector and on the relative excitation of the modes (see
\cite{Berti2006,Berti2007c} for detailed studies). Quantifying the excitation
of quasinormal modes for generic initial data is a formidable technical
problem \cite{Leaver1986,Berti2006a}, but numerical merger simulations allow
sensible estimates of the relative quasinormal mode amplitudes
\cite{Buonanno2006,Berti2007}. The relative mode amplitude depends on the
binary parameters, and therefore, in principle, the measurement of a
multi-mode signal with single or (better) multiple GW detectors can be used to
measure the binary mass ratio or the inclination of the final BH spin with
respect to the line of sight.

We also remark that the increased signal-to-noise ratio coming from
the merger has been shown to {\em improve} the bounds on alternative theories
of gravity that could come from observations of the inspiral only. Keppel and
Ajith, for example, discussed this possibility in the context of massive
graviton theories, where the graviton mass would modify the dispersion
relation of GWs \cite{Keppel:2010qu}. Formulations of the
evolution equations in alternative theories of gravity are in their infancy
\cite{Salgado:2008xh,Paschalidis:2011ww} and the numerical exploration of
BH binaries in alternative theories of gravity has just started
\cite{Healy:2011ef}. This is an interesting line of
research, and in the near future we may have more concrete ways to quantify
deviations from GR in strong-field mergers.

\vspace{3pt}
\noindent
{\bf \em GW template banks:} The key challenge in GW data analysis is to
accurately predict and isolate the features of gravitational waveforms in the
data stream from GW detectors. This is necessary in order to (i) detect the
presence of a GW signal of astrophysical origin, and (ii) determine the
parameters of the emitting source.  These goals can be achieved by matched
filtering, i.e.~by cross-correlating the data stream $s$ (composed of
detector noise $n$ plus a potential GW signal $h$) with a bank of theoretical
waveform templates $h_{\lambda_i}$, where the indices $\lambda_i$
$(i=1,\ldots,P)$ denote the $P$ intrinsic and extrinsic source parameters.
For this purpose it is imperative to construct a large set of accurate
numerical waveforms which cover the relevant parameter space. This task far
exceeds the capacity of purely numerical methods. If we assume a
seven-dimensional parameter space -- one mass ratio and three spin parameters
per hole -- a mere ten waveforms per parameter would correspond to $10^7$
waveforms. Furthermore, using binary waveform models including the long
inspiral phase plus merger optimizes the signal-to-noise ratio
in GW observations, and the inclusion of merger and ringdown has been shown
to provide significant improvements in source localization and distance
calculation \cite{McWilliams:2011zs}.
Because current numerical simulations cover only about 30 cycles
of the inspiral, the generation of complete waveforms requires the
combined use of PN and numerical methods. The construction
of such GW template banks currently proceeds along either of the following
two paths.

\noindent
(i) The {\em effective-one-body} (EOB) approach \cite{Buonanno1999,
Buonanno2000} maps the dynamics of the two-body problem in GR into the
motion of a particle in an effective metric. The components of the effective
metric are currently determined to 3PN order. The EOB method improves upon
this model by using additional {\em pseudo-PN} terms of higher
order\footnote{Comparing an EOB model without such additional calibration
against numerical relativity waveforms has been found to result in an
accumulated phase difference at merger of 3.6~rad for nonspinning binaries
with mass ratios $q$ between $1$ and $6$ \cite{Pan2011}.}  which are not
derived from PN expressions, but calibrated via comparison with numerical
results \cite{Pan2011,Buonanno2007,Damour2007a,Damour2008,
Pan2009,Taracchini:2012ig}.
Further improvements come from using a resummed version of the PN expanded
results and from modeling nonadiabatic effects in the inspiral: see
e.g.~Sec.~IV in \cite{Damour2010}. The inspiral-plunge waveform resulting from
this construction is then matched to a merger-ringdown signal composed of a
superposition of quasinormal oscillation modes of a Kerr BH. The total number
of free parameters varies between the individual EOB models currently
investigated, but in all cases, the general strategy is to calibrate these
parameters by comparison with a finite number of numerical
simulations.
% Furthermore, the inclusion of higher-order multipoles in the
%waveform models improves the measurement of BH binary parameters
%\cite{Lang2011} and the EOB model has recently been extended in this direction
%\cite{Pan2011}.

\noindent
(ii) {\em Phenomenological waveform templates} are based on {\em hybrid
  waveforms}: the early inspiral is modeled by PN techniques, and the
resulting signal is matched (within a specified window) to a numerical
waveform describing the last orbits, merger and ringdown.  The resulting set
of hybrid waveforms are then approximated by a model containing a number of
phenomenological parameters which are mapped to the physical parameters of the
binaries, such as the mass ratio $q$.
The phenomenological models initially
developed for nonspinning binaries in \cite{Ajith2007,Ajith2007a,Ajith2007b},
and then extended to the case of spins aligned with the orbital angular
momentum \cite{Ajith2009}, are based on hybrid waveforms matched in the {\it
time domain}, whereas the more recent study by Santamaria {\em et al.}
\cite{Santamaria2010} performs the matching in the frequency domain.
In spite of such differences in their construction, the final result
of all these phenomenological models are closed-form analytic expressions
for the waveforms in the frequency domain. An exploration of phenomenological
models for equal-mass binaries with spin precession is given in
\cite{Sturani2010a}.
%%
%\eb{Uli, the Ajith models are also frequency-domain in the end. Are you sure
%  about this statement? If it's true, I guess it refers only to the
%  calibration, but it gives the impression that implementing the Ajith models
%  needs time evolutions, and it doesn't.}
%\us{I have added a sentence to clarify this. I indeed only meant the hybrid
%construction. Still to do: Add the Sturani articles which I so far did not
%quite understand...}

Both types of template banks are employed in the analysis of GW data. The
recent search for GWs from binary BH inspiral, merger and ringdown
\cite{Abadie2011} used phenomenological models for injection, and an EOB
model for injection and for the search templates. This search constrained the
rate of mergers of binaries with individual masses in the range
$[19, 28]~M_{\odot}$ to be no more than $2.0~{\rm Mpc}^{-3}{\rm
  Myr}^{-1}$ at 90\% confidence.  Numerical waveforms have also been used
inside the GW community-wide {\em Ninja} project \cite{Ninja} to study the
sensitivity of existing GW search algorithms used in the analysis of
observational data.  For this purpose, numerical relativity waveforms were
injected into a simulated data stream designed to mimic ground-based detector
noise.  The algorithms detected a significant fraction of the injections, but
likely require further development, in particular for the purpose of measuring
source parameters \cite{Aylott2009,Aylott2009a,Ajith:2012tt}.
A further community wide
effort, the {\em NRAR} project \cite{NRAR}, is dedicated to a systematic
exploration of the complete BH binary parameter space to develop
optimally-calibrated template families to be used in GW data analysis.

%\noindent
%\us{Move some of this to Sec.~III.}
%\eb{I kept what I think is REALLY essential here...}
One of the most exciting prospects of GW detection is the idea of
looking for electromagnetic counterparts to the GW signal.  While a {\it
  network} of Earth-based detectors is required to localize a GW source by
triangulation (see e.g.~\cite{Cutler1994}), a single space-based detectors
like LISA may suffice to localize low-redshift sources and determine their
luminosity distance. 
%Binary BH waveform models including merger yield a larger signal-to-noise
%ratio than models including inspiral only. Furthermore,
The inclusion of merger and/or of higher multipolar components of the
radiation in the waveform models has been shown to provide significant
improvements in source localization and distance determination
\cite{Lang2011,McWilliams:2011zs}.
For this reason it is very important to construct
phenomenological models including higher multipoles of the full
inspiral/merger/ringdown waveform. The EOB model has recently been extended in
this direction \cite{Pan2011}.
%
%The waveform modulations induced by spin precession have also been shown to
%yield significant improvements in parameter estimation. If spin alignment is
%indeed efficient, as argued above, the spin precession-induced modulations
%will be relatively small. This may have a negative impact on source
%localization and distance determination, as shown by Lang {\it et al.} for
%inspiral-only waveforms \cite{Lang2011}. More systematic studies of
%``complete'' inspiral/merger/ringdown waveforms including spin precession are
%needed to realistically estimate the potential of future GW detectors to work
%in conjunction with traditional electromagnetic observatories.

\vspace{3pt}
\noindent
{\bf \em Accuracy requirements:} Clearly, the two approaches for template
construction mentioned above require accurate numerical waveforms. Quantifying
these accuracy requirements is a nontrivial task. Numerical uncertainties due
to finite resolution are directly tested by {\em convergence analysis}, and
the error incurred by extracting GWs at finite radius can be estimated using
extrapolation of results from various radii to infinity \cite{Boyle2009a}.
The uncertainties in phase and amplitude thus derived, however, depend on the
alignment of the waveforms in phase and time; see e.g.~Fig.~2 in
\cite{Sperhake2010a}. Because offsets in phase and time are free parameters in
GW data analysis, it is more convenient to measure accuracy and agreement of
different waveforms in terms of quantities which take such alignment into
account by construction.
Such a measure is obtained from the inner product between two waveforms $h(t)$
and $g(t)$ used in the theory of parameter estimation
\cite{Finn1993,Cutler1994}:
\begin{equation}
  \langle h,g \rangle \equiv 4 \mathbf{Re} \int_0^{\infty}
      \frac{\tilde{h}(f) \tilde{g}^*(f)}{S_N(f)} df,
      \label{eq:innerprod}
\end{equation}
where the tilde and asterisk denote Fourier transform and complex conjugate,
respectively, and $S_N(f)$ is the one-sided power spectral density of the
detector strain noise \cite{LIGOnoisecurves}. In practice, this inner product
is to be understood as maximized over constant offsets in time and phase
($\Delta t_0$ and $\Delta \phi_0$) between the two waveforms\footnote{For
detection of GW events, one often considers the {\em effectualness} of
waveform models (as opposed to {\em faithfulness}) and in those cases
also maximizes the inner product over the source parameters.}.  The
signal-to-noise ratio that would be obtained for a physical signal $h_e$ and a
model waveform $h_m$ is then given by $\rho_m = \langle h_e | h_m \rangle / ||
h_m ||$. It is related to the optimal signal-to-noise ratio $\rho$ for a
perfect model waveform by the {\em mismatch} $\mathcal{M}$ \cite{Cutler1994}:
\begin{equation}
  \rho_m = (1-\mathcal{M}) \rho = (1-\mathcal{M}) \langle h_e | h_e \rangle
      / || h_e ||.
  \label{eq:mm1}
\end{equation}
To leading order in $||\delta h||$, this definition of the mismatch implies
\cite{Lindblom2010,MacDonald2011}
\begin{equation}
  \mathcal{M} = \frac{\langle \delta h | \delta h \rangle
      - \langle \delta h_{||} | \delta h_{||} \rangle}
      {2\langle h_e | h_e \rangle}
      \le \frac{\langle \delta h| \delta h \rangle}
      {2\langle h_e | h_e \rangle},
      \label{eq:mm2}
\end{equation}
where $\delta h_{||} = \langle \delta h | h_e \rangle / \langle h_e | h_e
\rangle$ is that part of the waveform error parallel to $h_e$.

Based on these definitions, accuracy requirements are obtained as follows.
Two waveforms differing by $\delta h$ will be indistinguishable for a
detector if $\langle \delta h | \delta h \rangle < 1$ \cite{Lindblom2010,
Damour2010},
%\vc{I am quite positive this previous statement cannot be correct: I
%should be free to specifiy the amplitude of my event without changing
%the accuracy requirements, but the formula does not allow that. Can you
%tell me exactly where this formula appears?}
such that acceptable waveform models to be used in parameter estimation must
lie within a sphere of unit radius of the exact waveform.  For the purpose of
event detection, we note that the fractional loss in detection of GW signals
due to imperfect templates is proportional to the third power of the reduction
in the SNR, i.e.~$\approx 3\mathcal{M}$. A value of $10\%$ is typically deemed
acceptable, corresponding to a mismatch $\mathcal{M}=3\% \ldots 3.5\%$.  For
detection efficiency, there arises, however, a complication due to the
discrete nature of the template bank. As discussed in detail in Sec.~II B of
\cite{Lindblom2010}, the effective mismatch in a real GW search is given by a
contribution due to the actual modeling procedure, plus a term which accounts
for the discrete spacing in the parameter space: see their Fig.~3 and
Eq.~(17). Template banks adopted in LIGO searches use 0.3 for the latter
\cite{Abadie2011,Abbott2007b} and, in consequence, Lindblom {\em at al.}
\cite{Lindblom2010} recommend a maximum mismatch
$\mathcal{M}_{\rm max}=0.005$. In terms of the right-hand side of
Eq.~(\ref{eq:mm2}), dubbed {\em inaccuracy function} in
Ref.~\cite{Damour2010}, we summarize the accuracy requirements as
(cf.~Eq.~(10) in Ref.~\cite{MacDonald2011})
\begin{eqnarray}
  \frac{||\delta h||}{||h||} < \left\{
      \begin{array}{ll}
        1/\rho & {\rm for\,\,\,parameter\,\,\,estimation,} \\
        \sqrt{2\mathcal{M}_{\rm max}} & {\rm for\,\,\,detection.}
      \end{array}
      \right.
\end{eqnarray}
%
%\vc{Following is too lengthy in my opinion}
%\eb{May be but it's very important. I tried to shorten a bit.}
Accuracy measurements employing the mismatch or $||\delta h||$ have been used
to study numerical, semi-analytic and hybrid waveforms. According to the
Samurai project \cite{Hannam2009}, relative mismatches of the $l=m=2$ mode
between a variety of purely numerical waveforms for binaries with total
mass\footnote{For significantly lower masses, the last 10 orbits and merger
  signal lie outside the maximum sensitivity range of the current generation
  of ground-based GW detectors.}  $M\ge 60~M_{\odot}$ are better than
$10^{-3}$.
Hannam {\em et al.}  \cite{Hannam2010a} found errors in the
construction of hybrid waveforms to be dominated by PN
contributions. Numerical waveforms should be long enough to enable matching at
lower frequencies, where PN approximations are more accurate: $\sim 3$ orbits
before merger for the equal-mass, nonspinning case, and $\sim 10$ orbits for
binaries with spins $\chi=0.5$. MacDonald {\em et al.}  \cite{MacDonald2011}
report significantly more demanding requirements of $\sim 30$ orbits for
nonspinning, equal-mass binaries. Similarly, Boyle \cite{Boyle2011} concludes
that longer NR waveforms or more accurate PN predictions will be required.
The discrepancy largely arises from the use
of more stringent accuracy thresholds: $\mathcal{M}_{\rm max}=0.005$
in \cite{MacDonald2011}
%i.e.~$||\delta h||/||h||=0.01$ 
(instead of $\mathcal{M}_{\rm max}=0.03$
%i.e.~$||\delta h||/||h|| = 0.245$ 
in \cite{Hannam2010a}) for detection,
%taking into account the discrete spacing of template banks discussed above,
and $\rho=40$ for parameter estimation. Most recently, Ohme {\em et al.}
\cite{Ohme2011} have argued that maximising the match over intrinsic
source parameters as well as phase and time shifts may reduce length
requirements on numerical waveforms.
%Figs.~4 and 7 in \cite{MacDonald2011} illustrate the sensitivity of $||\delta
%h||/||h||$ on numerical and hybridization error in this regime.

%Using $\mathcal{M}_{\rm max}=0.03$, Hannam {\em et al.}
%\cite{Hannam2010a} have studied length requirements for the construction of
%hybrid waveforms. They find errors to be dominated by PN contributions, which
%imposes a length requirement on numerical waveforms to enable matching at
%lower frequencies where PN approximations are more accurate: about three
%orbits before merger for the equal-mass, nonspinning case, and ten orbits for
%binaries with moderate spins $\chi=0.5$. In comparison, MacDonald {\em et al.}
%\cite{MacDonald2011} report more demanding requirements of about
%30 orbits. The discrepancy largely arises from the use of different accuracy
%thresholds $\mathcal{M}_{\rm max}=0.005$, i.e.~$||\delta h||/||h||=0.01$
%(instead of $0.03$, i.e.~$||\delta h||/||h|| = 0.245$ in
%Ref.~\cite{Hannam2010a}) for detection, taking into account the discrete
%spacing of template banks discussed above, and $\rho=40$ for parameter
%estimation; see Figs.~(4), (7) in \cite{MacDonald2011} for the sensitivity of
%$||\delta h||/||h||$ on numerical and hybridization error in this regime.
%
A comparison of phenomenological and EOB template banks based on the
inaccuracy function has been performed by Damour {\em et al.}
\cite{Damour2010}.  For advanced ground-based detectors, they conclude that
current phenomenological models deviate from the EOB (considered as a target
model) beyond acceptable thresholds over a wide range of the parameter space,
for both, detection and parameter estimation.

Further reading on BH simulations in the context of GW
physics is available in \cite{Pretorius2007a,Hannam2009a,
Centrella2010,Hinder2010}.

%The signal-to-noise ratio is defined in terms of the convolution of a template
%$h$ with the data $s$ weighted by the detector specific one-sided
%noise power spectral density $S_N$ \cite{LIGOnoisecurves}
%%
%\begin{equation}
%  \rho \equiv \frac{\langle s,h \rangle}{{\rm rms}\langle n, h \rangle},
%  \qquad
%  \langle u,v \rangle \equiv 4 \mathbf{Re} \int_0^{\infty}
%      \frac{\tilde{u}(f) \tilde{v}^*(f)}{S_N(f)} df,
%\end{equation}
%%
%where the tilde and asterisk denote Fourier transform and complex conjugate,
%respectively, and
%``rms'' denotes the root-mean-square over all realizations of the detector
%noise.
%GW detection is determined by $\rho$ exceeding some predetermined threshold,
%while the source parameters $\lambda_i$ are identified with that particular
%template waveform which maximizes the signal-to-noise ratio.

%=============================================================================
\section{Numerical relativity and astrophysics}
\label{sec:astro}

%\eb{Topics: 1) astro$\rightarrow$ NR: mass/spin measurements, and possible
%  measurements of $Q$ from dynamics near the galactic center; event rates on
%  Earth and in space (and progress in SMBH binary observations); IMRIs for
%  Earth- and space-based detectors.}

Astrophysical observations have provided the strongest evidence to date that
BHs exist and play an important role in many physical processes in our
Universe.  Astrophysical observations of X-ray binaries or quasars, as
mentioned in Sec.~\ref{sec:intro}, have a good deal to tell us about BHs. In
recent years, numerical relativity simulations of BHs have also deepened our
insight into a variety of astrophysical systems.  We will first summarize what
we have learned from ``traditional'' electromagnetic observations about the
populations of BHs we can expect to exist in the Universe. Then we will
discuss some astrophysical implications of numerical relativity.

%There are several excellent reviews on measuring BH masses, spins and
%(possibly) providing evidence of an event horizon by ``traditional''
%electromagnetic astronomy \cite{Narayan:2005ie} and on using electromagnetic
%observations of BHs and neutron stars to probe strong-field gravity
%\cite{Psaltis:2008bb}. Here we will focus on the interplay between numerical
%relativity and BH astrophysics, and in particular on the following questions:
%(1) What do astrophysical observations tell us about the masses and spins of
%BH binaries, and on binary merger event rates? (2) Vice versa, what are the
%implications of numerical simulations for present and future observations of
%astrophysical BHs?\vc{Following 3-4 pages are too lengthy. I would suggest to reduce them in half or more}.

\vspace{3pt}
\noindent
{\bf \em Expected BH populations:} We have already mentioned the two main
classes of BHs identified by astrophysical observations.  (i) Solar-mass BHs
with $M\sim 5-20~M_\odot$ are usually found in X-ray binaries
\cite{Remillard:2006fc} and are formed as the end-product of the evolution of
massive stars.  (ii) SMBHs with $M\sim
10^6-10^{9.5}~M_\odot$ are believed to reside in most Active Galactic Nuclei
(AGN) \cite{Melia:2007vt}. The assembly of these SMBHs is likely to result
from a combination of BH mergers and accretion of surrounding matter over
cosmological timescales. Evidence for a third class of BHs
with $M\sim 10^2-10^5~M_{\odot}$ (IMBHs) is tentative at present
\cite{Miller:2003sc,Hurley:2007as,Berghea:2008rc}.

GW frequencies are inversely proportional to the system's total mass, and
therefore BH masses play a crucial role in binary BH detectability.
Earth-based detectors, such as LIGO and Virgo, have an optimal sensitivity
band corresponding to stellar-mass BHs and IMBHs, while the planned
space-based detector LISA is most sensitive to high-mass IMBHs and SMBHs.  As
discussed in the previous section, the expected GW pattern from BH binaries
depends on the masses and spins of the binary members. We shall further see
below that the gravitational recoil imparted upon the remnants of coalescing
binaries strongly depends on spins and mass ratio. So, what do ``traditional''
electromagnetic observations tell us about BH masses and spins?

%
%For this reason, we will briefly review the present understanding of BH mass
%and spin distributions. Somewhat artificially, we will draw a distinction
%between ``low-mass binaries'' of interest for Earth-based GW detectors and
%``high-mass binaries'' to be observed by space-based GW detectors (see
%\cite{Berti2009} for more details).
%
%\noindent
%{\bf \em Low-mass black hole binaries:} 
%
%The most accurate mass measurements for stellar-mass BH candidates are made by
%looking at how the unseen BH affects the orbit of a companion star.  In the
%case of BH X-ray binaries it is relatively easy to measure the binary orbital
%frequency $\omega$ and the maximum line-of-sight Doppler velocity $K_c=v\sin
%\iota$ of the companion star. From these quantities one can compute a ``mass
%function'' $f(M)$ that provides a strict {\em lower bound} on $M$. Since NSs
%must be less massive than $\sim 3M_\odot$ \cite{Rhoades:1974fn}, all X-ray
%binaries with $f(M)\gtrsim 3M_\odot$ should contain a BH.  Remillard and
%McClintock \cite{Remillard:2006fc} review observations of 20 X-ray binaries
%with dynamically confirmed BHs and discuss proposed methods for measuring
%their spins.
%The heaviest stellar-mass BH candidate to date is IC 10 X-1, with $M\geq
%23.1\pm 2.1M_\odot$ (and possibly even larger) \cite{Silverman:2008ss}. This
%system is particularly interesting for GW detection: it should become a close
%double BH binary with coalescence time of $\sim 2-3$~Gyr, implying that BH
%binary merger rates for Earth-based detectors may be higher than anticipated
%\cite{Bulik2011}.

Current spin estimates for accreting, stellar mass BHs are usually obtained by
modeling the shape of their X-ray spectrum or by analyzing the skew in
Fe~K$\alpha$ emission lines. The resulting estimates are all, to some extent,
model-dependent, but frequently yield moderate values $\chi \approx 0.1 \ldots
0.8$ and, for some cases, values close to the Kerr limit $\chi=1$;
see e.g.~\cite{Berti2009,McClintock:2011zq}
%\cite{Brenneman:2009hs,McClintock:2009as,Miller:2009hh}
and references therein.  Theoretical arguments and observations suggest that
stellar-mass BHs in binaries {\it retain the spin they had at birth}: neither
accretion nor angular momentum extraction are likely to change significantly
their mass or spin \cite{King:1999aq,Belczynski:2007xg}.  For SMBHs, spin
estimates relying on the shape of the Fe~K$\alpha$ line yield values ranging
from moderate $\chi \sim 0.6 \ldots 0.8$ (e.g.~\cite{Miniutti:2009dg,
  Schmoll:2009gq,Gallo:2010mm} to near-critical spins
(e.g.~\cite{Fabian:2005qj,Brenneman:2006hw,Fabian:2009yc,Brenneman:2011wz}).
SMBHs are expected to grow by a combination of mergers and accretion.  Their
spin depends sensitively on the details of these processes and of their growth
\cite{Gammie2003,Hughes2003,Berti2008}.  SMBH assembly via BH mergers does not
appear to be able to account for the large observed spin values; this may
indicate that accretion dominates over mergers \cite{Volonteri2004}. In any
case, BH spins encode the history of their formation, and it would be
extremely useful to have detailed knowledge of the BH spin distribution
function.  Ultimately, such a detailed census of BH parameters is one of the
key targets of future GW observations \cite{Mandel:2010xq,Sesana:2010wy}.  For
the purpose of numerical modeling of BHs, present measurements indicate that
all spin magnitudes ($0\leq \chi\leq 1$) should be covered.
%Few dependable and
%accurate measurements are available (see e.g. Table 3 of \cite{Berti2009} for
%references), but theoretical arguments and observations suggest that
%stellar-mass BHs in binaries {\it retain the spin they had at birth}: neither
%accretion nor angular momentum extraction are likely to change significantly
%their mass or spin \cite{King:1999aq,Belczynski:2007xg}.
%  Precession should
%only marginally impact inspiral GW detection rates, but it will be significant
%for parameter estimation
%\cite{O'Shaughnessy:2005qc,Belczynski:2007xg,Grandclement:2003ck}. Therefore,
%the accurate modelling of precessing binaries may allow accurate measurements
%of the spins of stellar-mass BHs,

We have already noted that stellar-mass BHs typically have masses in the range
$[5, 20]~M_{\odot}$. Unfortunately, there are no confirmed observations
of stellar-mass {\em binary} BH candidates, so estimates of mass ratios must
rely on theoretical models. For stellar-mass binaries, population synthesis
codes suggest that $q$ should always be quite close to unity
\cite{Belczynski:2007xg}.  Measurements of SMBH masses are obtained by
observations of stellar motion near galactic centers, as mentioned in
Sec.~\ref{sec:intro}, as well as motion of gas discs \cite{Herrnstein:1999cw},
applications of the virial theorem to the velocity dispersion of stars
\cite{Kormendy1995} and reverberation mappings applied to more distant AGNs
\cite{Gebhardt:2000sb}.  An exhaustive list of galaxies with SMBH mass
measurements in the range $M\sim 10^5-10^9~M_{\odot}$ is presented by Graham
{\em et al.}  \cite{Graham:2008uh,Graham:2010nb}.
%which lie in the range $M\sim 10^5-10^9~M_{\odot}$.
The impact of different SMBH assembly models on the mass and mass ratio
distribution of detectable binaries has been discussed by various authors. The
general consensus is that mass ratios $q\lesssim 1/10$ (and down to $q\approx
10^{-4}$) should be common
\cite{Sesana2007,Koushiappas:2005qz,Gergely:2008mt}.  In contrast with the
case of stellar-mass BHs, there is by now some observational evidence for SMBH
binaries \cite{Komossa:2003wz,Bianchi:2008un,Valtonen:2008tx,
  Rodriguez:2009ax,Civano:2010es}.
%The earliest convincing observational case for a ``tight'' SMBH binary was the
%ultraluminous infrared galaxy NGC 6240 \cite{Komossa:2002tn}, containing two
%active SMBHs separated by a relatively short projected distance $\sim 1$~kpc.
%Since then, many more promising SMBH binary candidates have been identified:
%four of these are at separations $\sim 1$~kpc
%\cite{Komossa:2002tn,Ballo:2003ww,Bianchi:2008un,Civano:2010es},
%one (0402+379) is at $\sim 10$~pc \cite{Rodriguez:2006th,Rodriguez:2009ax} and
%one system (OJ287) has been interpreted as a close binary at separation $\sim
%0.1$~pc \cite{Valtonen:2008tx}.
These observations are not sufficient to
tightly constrain SMBH merger rates, but they are at least broadly consistent
with merger-tree models predicting tens to hundreds of events during the
typical lifetime of a space-based GW detector \cite{Volonteri2009}.

%The relative importance of merger with respect to inspiral waves decreases for
%extreme mass ratios (see e.g.~\cite{Berti2007}).
A more speculative kind of source for Earth-based detectors consists of the
intermediate mass ratio inspiral of a compact object (neutron star or BH) into
an IMBH.
% The relatively low energy content in merger waves is
%compensated, in this case, by the fact that the ringdown frequency is close to
%the minimum of the Advanced LIGO noise power spectral density
%\cite{Mandel:2007hi}.
Another promising source for advanced Earth-based interferometers (albeit with
highly uncertain event rates) are IMBH-IMBH mergers.  The initial inspiral
of these binaries could be detected via space-based interferometers, while the
ringdown phase is in the optimal bandwidth for second- and third-generation
detectors such as Advanced LIGO and ET, that could therefore be used for
``follow-up'' ringdown searches \cite{Fregeau:2006yz,AmaroSeoane:2009ui}.

As in the case of BH spins, the observations imply that numerical relativity
needs to cover a wide range in $q$. This can be done most efficiently by
bridging the gap between numerical studies and the perturbative modeling of
extreme-mass-ratio binaries
(e.g.~\cite{Hinderer2008,Barack2009,Poisson2011}).

From this discussion, it is clear that a detailed knowledge
of the spin evolution as well as the generation of gravitational recoil
in BH binary mergers is important for understanding
the cosmological evolution of SMBHs over cosmological times. We will
discuss these effects in turn.

\vspace{3pt}
\noindent
{\bf \em Black-hole spins resulting from mergers:} Prior to the 2005 numerical
relativity breakthroughs, it was known that ``minor mergers'' ($q\lesssim 10$)
of a large, rotating BH with an isotropic distribution of small objects would
tend to spin down the hole \cite{Hughes:2002ei}. Numerical merger simulations
showed that the merger of comparable-mass, nonspinning BHs leads to a final
Kerr BH with spin parameter $a/M=0.69$. The simulations were followed by the
development of several models to predict the spin of merger remnants as a
function of the binary parameters for generic mass ratios and spins.

The first studies focussed on performing numerical evolutions of the last few
orbits of BH binaries, and used the results to calibrate formulae that map the
initial binary parameters to values for the spin of the merged hole
\cite{Campanelli2006d,Berti2007,
  Marronetti2007a,Rezzolla2007,Rezzolla2007a,Rezzolla2007b}.  It became clear,
however, that the binary inspiral up to the last orbits has the potential to
significantly affect the spin distribution (e.g.~\cite{Schnittman2004}) and
therefore should be included, for example via PN modeling, in the derivation
of maps from initial parameters to the merger remnant properties
\cite{Tichy2008,Barausse2009,Lousto2009,Lousto2009a,Kesden2010}.  Predictions
for the final spin based on the extrapolation of test-particle calculations to
finite mass ratios have been developed in \cite{Buonanno2007,Kesden2008}, and
show remarkably good agreement with numerical results in the comparable-mass
regime.  A comprehensive review of all spin formulae is beyond the scope of
this article, but we refer the reader to the review in \cite{Rezzolla2008} and
the discussion in Sec.~V of \cite{Kesden2010}.
% These studies provided
%predictions for the spin resulting from individual merger events
%\cite{Buonanno2007,Barausse2009} and they showed that PN effects
%may alter the distribution of BH spins in the presence of partial spin
%alignment at large separations \cite{Kesden2010}.
%\us{Collecting some results here:}
If we assume an ensemble of BH binaries with initially randomly
oriented spins, the final spin distribution is peaked around
$\chi_f\approx 0.7$ \cite{Berti2008,Tichy2008,Lousto2009a}.
%report a universal distribution highly peaked
%around $\chi_f=0.73$ and a $25^\circ$ misalignment relative to the
%final orbital angular momentum when starting simulations from $50~M$.
Campanelli {\em et al.} \cite{Campanelli2006d,Campanelli2007} reported spin
flips by $34-103^{\circ}$ with respect to the initial spin direction of the
larger hole; spin flips of this magnitude could provide an explanation for
X-shaped radio sources \cite{Merritt2002}.  Kesden {\em et al.}
\cite{Kesden2010} demonstrated that spin precession over the course of a long
inspiral of thousands of orbits tends to align (antialign) the binary BH spins
with each other if the spin of the more massive BH is initially partially
aligned (antialigned) with the orbital angular momentum, thus increasing
(decreasing) the average final spin. Spin precession is stronger for
comparable masses, and it could produce significant spin alignment before
merger for both SMBHs and stellar-mass BH binaries.
%\us{End of collected results.}

\vspace{3pt}
\noindent
{\bf \em Gravitational recoil:} One of the most spectacular results obtained
from numerical BH binary simulations is the quantitative prediction of the
magnitude of {\em gravitational recoils} (or {\em kicks}). In the 1960s it was
realized that the emission of linear momentum via GWs must impart a kick on
the source due to momentum conservation \cite{Bonnor1961,Peres1962}. In the
1980s, the kick generated in compact binary inspirals and plunges was studied
in the framework of PN theory and BH perturbation theory
\cite{Fitchett1983,Fitchett1984,Nakamura1984}. However, the astrophysical
relevance of the gravitational recoil following BH binary mergers remained an
open question until the recent numerical relativity breakthroughs.

For the case of an equal-mass, nonspinning BH binary, the net linear momentum
emitted in GWs vanishes due to symmetry.  Nonzero recoils are therefore only
generated in systems where this symmetry is broken through (i) a mass ratio
$q<1$ (equivalently, a symmetric mass-ratio parameter $\eta\equiv
q/(1+q)^2<1/4$) or (ii) nonvanishing spins.

For zero spins, early numerical studies
%Performing a least-square fit formula to data 
in the range $q=[1,1/4]$
%Gonz{\'a}lez
%{\em et al.} \cite{Gonzalez2007} 
found that the kick velocity is well approximated by
\cite{Herrmann2006,Baker2006b,Gonzalez2007}
\begin{equation}
  v_{\rm kick} = 1.2\times 10^4 \eta^2\sqrt{1-4\eta}(1-0.93\eta)~{\rm km/s}.
      \label{eq:nonspinningkick}
\end{equation}
This result (whose functional form is inspired by Fitchett's analytical work
\cite{Fitchett1983})
%is in excellent agreement with the nonspinning limit of Eq.~(4) in Baker {\em
%et al.} \cite{Baker2008}, and it
was confirmed by simulations for smaller mass ratios ($q=1/10$,
\cite{Gonzalez2008}) and by analytical work \cite{LeTiec2010}.  The maximal
recoil obtained from Eq.~(\ref{eq:nonspinningkick}) is $v_{\rm max}=175.2\pm
11~{\rm km/s}$ for $q=0.36\pm 0.03$.  Small orbital eccentricity $e\le0.1$
should introduce corrections proportional to $e$ \cite{Sopuerta2007}.

Spinning binaries are characterized by seven free parameters (the mass ratio
plus three components for each BH spin), and numerical studies inevitably
focussed on subsets of the parameter space. It soon became clear that the
spin interaction dominates over mass ratio effects. The first studies for
equal-mass binaries with spins parallel to the orbital angular momentum
revealed kicks of several hundreds km/s, with an extrapolated maximum of $\sim
500~{\rm km/s}$ for extremal spin magnitudes \cite{Herrmann2007,Koppitz2007}.
Shortly thereafter, the discovery of the so-called {\em superkicks} marked one
of the most surprising outcomes of numerical relativity: binaries for which
the spins are perpendicular to the orbital angular momentum and anti-aligned
with each other can generate kicks of {\em thousands} of km/s, with an
extrapolated maximum $v_{\rm max}\sim 4000~{\rm km/s}$ for near-extremal spin
magnitudes \cite{Campanelli2007,Gonzalez2007a,Campanelli2007a}.
Most recently, BH binaries with spins partially aligned with the orbital
angular momentum but whose spin projections into the orbital plane
still correspond to the superkick configurations have been found
to result in even larger maximum recoils of up to $\sim 4\,900$~km/s
\cite{Lousto:2011kp,Lousto:2012su}, so-called ``hang-up kicks''.

Many galaxies harbor SMBHs at their centers, and galaxy mergers should
generally lead to the merger of their central BHs \cite{Begelman:1980vb}.
Typical escape velocities range from $\sim 10~{\rm km/s}$ for dwarf galaxies
up to $\sim 1000~{\rm km/s}$ for giant elliptic galaxies \cite{Merritt2004}.
Large kicks would displace or eject the merged hole from its host, with
possibly observable consequences: a softening of the stellar density gradient
in the galactic nucleus, off-center radio-loud active galactic nuclei,
off-nuclear X-ray sources in nearby galaxies
%the ejection of BHs from galaxies and their dark-matter halo progenitors,
and the generation of electromagnetic signals via interaction of the BH with
its gaseous environment
\cite{BoylanKolchin2004,Haiman2004,Madau2004,Merritt2004,
  Libeskind2006,Loeb2007,Komossa2008b,Lippai2008,Guedes2009,
  Chang2010,Guedes2011}.  BH ejection represents a potential obstacle for BH
growth via merger, and thus puts constraints on merger-history models, which
must be able to explain the assembly of SMBHs by redshifts $z\gtrsim 6$
\cite{Haiman2004,Li2007,Volonteri2010}. Observed redshifts of broad-line
relative to narrow-line regions in quasar spectra may be interpreted as a
smoking gun of BH ejection due to gravitational recoil
\cite{Komossa2008,Civano:2010es,Robinson2010}, but alternative interpretations
(such as a BH binary, or superposed emission regions from two interacting
galaxies) are possible \cite{Bonning2007,Bogdanovic:2008uz,
  Shields2008,Vivek2009}.

The interest in astrophysical consequences of large recoils led to
numerical studies of the BH parameter space
\cite{Tichy2007,Baker2007,
  Bruegmann2007,Lousto2007,Pollney2007,Baker2008,Dain2008,Lousto2008,
  Lousto2010a,Zlochower2010}. Phenomenological kick formulas inspired by PN
studies \cite{Kidder1995} and similar to Eq.~(\ref{eq:nonspinningkick}) were
proposed to map the input parameters of a given binary configuration to the
final kick velocity. Available numerical results span mass ratios in the
range $q\ge 1/10$ and spin magnitudes $|\boldsymbol{\chi}|\le 0.9$, and are
well described by Eq.~(2) of Ref.~\cite{Campanelli2007}
%well described by the formula \cite{Campanelli2007}
%%
%\begin{equation}
%  \mathbf{V} = v_m \hat{\mathbf{e}}_1 + v_{\bot}(\cos~\xi~\hat{\mathbf{e}}_2
%      + \sin~\xi~\hat{\mathbf{e}}_2) + v_{||} \hat{\mathbf{e}}_z,
%      \label{eq:spinningkick}
%\end{equation}
%%
%\vc{The formula above is very unclear--do we need to show this?--also, do we say what $\mathbf{V}$ or $\boldsymbol{\chi}$ are?}
%where $v_m=A \eta^2 \frac{1-q}{1+q} (1+B\eta)$, $v_{\bot}=H\eta^2
%\mathbf{\Delta}^{||}\cdot \hat{\mathbf{e}}_z$, $v_{||}=K\eta^2
%\cos(\Theta-\Theta_0) |\mathbf{\Delta}^{\bot}|$, $\hat{\mathbf{e}}_{1,2,z}$ is
%an orthonormal basis with $\hat{\mathbf{e}}_z$ parallel to the orbital angular
%momentum $\mathbf{L}$, and $\mathbf{\Delta}^{||,\bot}$ are the components of
%$\mathbf{\Delta}= (q\boldsymbol{\chi}_2-\boldsymbol{\chi}_1)/(1+q)$ parallel
%and perpendicular to $\mathbf{L}$. $\Theta$ is the angle between
%$\mathbf{\Delta}^{\bot}$ and the infall direction at merger, and $\Theta_0$
%depends on the mass ratio but not on the spins \cite{Lousto2009}. In practice
%one is often interested in the maximal or average component $v_{||}$ and
%replaces $\cos~(\Theta - \Theta_0)$ accordingly. In addition to the
%coefficients $A=1.2\times 10^4~{\rm km/s}$, $B=-0.93$ from
%Eq.~(\ref{eq:nonspinningkick}), fitting available numerical data yields
%$H=(6.9\pm0.5)\times 10^3~{\rm km/s}$, $K=(6.0\pm 0.1)\times 10^4~{\rm km/s}$
%and $\xi \sim 145^{\circ}$ \cite{Campanelli2007a,Lousto2007}.
(see also \cite{Lousto2009,vanMeter2010}).
%The large
%value of $K$ in Eq.~(\ref{eq:spinningkick} mirrors the large magnitude of the
%superkicks pointing out of the orbital plane.

%\vc{Is the following paragraph really necessary?}
%\us{I will substantially shorten this, but I would like to mention at least
%the most important consequences of recoil for astrophysics.}
%%
%\us{let's see whether we have space to discuss the final parsec problem} 
%%
%\eb{Not important enough, it's generally considered to be solved}
%%

%The interest in astrophysical consequences of large recoils led to
%phenomenological studies of the {\em distribution} of kick velocities in the
%seven-dimensional space of binary parameters
%\cite{Schnittman2007,Kesden2010a}.
The apparent incompatibility between large
recoils and the existence of BHs at galactic centers can be resolved by
mechanisms that would {\it align} the individual BH spins with the orbital
angular momentum of the binary.  One such mechanism are gas torques in the
so-called ``wet'' (gas-rich) mergers, that would produce partial alignment
``early on'' in the inspiral phase \cite{Bogdanovic2007}. If partial alignment
in gas-rich mergers is the norm, as suggested also by the spin measurements
discussed above, PN spin effects will lead to {\it further} alignment of the
spins with the orbital angular momentum, significantly reducing the typical
values of recoil velocities \cite{Schnittman2004,Kesden2010a}.
This mechanism
has even been shown capable of efficient suppression of large recoil
velocities in the case of the above mentioned hang-up kicks
\cite{Berti:2012zp}.

\vspace{3pt}
\noindent
{\bf \em Merger simulations with matter and the Blandford-Znajek effect:} Many
numerical relativity groups are presently working on binary BH simulations in
the presence of matter. Most of this work is trying to understand the
signature of electromagnetic counterparts to binary BH mergers: for example,
such a signature could be produced by merger events \cite{Megevand:2009yx} or
recoiling BHs \cite{Anderson:2009fa} ``shocking'' the surrounding accretion
disks. The Georgia Tech group presented the first simulations in full GR of
equal-mass, spinning BHs merging in a gas cloud. They found that shocks,
accretion and relativistic beaming can produce electromagnetic signatures
correlated with GWs, especially when spins are aligned with the orbital axis
\cite{Bode:2009mt}. Later work by the group focussed on hot, radiatively
inefficient accretion flows. In this case, BH binaries exhibit a flare
followed by a sudden drop in luminosity associated with the merger, and
quasi-periodic oscillations correlated with the GWs during
inspiral \cite{Bogdanovic:2010he,Bode:2011tq}.  The Urbana group simulated
equal-mass, nonspinning BH binaries embedded in gas clouds under different
assumptions for the motion of the binary with respect to the gas. They found
evidence that the accretion rate and luminosity due to bremsstrahlung and
synchrotron emission would be enhanced with respect to a single BH of the same
mass as the binary, possibly being detectable by LSST
\cite{Farris:2009mt}.
%\cite{Etienne:2010ui} is work building infrastructure; \cite{Liu:2010mh} is
%relevant but has nothing to do with numerical relativity, so we can ignore
%it.
Recent work suggest that the circumbinary disk surrounding BH
binaries should not produce detectable electromagnetic counterparts
\cite{Bode:2011tq,Moesta:2009rr}. However, the {\em magnetic fields}
produced by the circumbinary disk could affect the binary dynamics
\cite{Noble:2012xz}, produce stronger electromagnetic counterparts
\cite{Giacomazzo:2012iv} and extract energy from the orbiting BHs,
which ultimately merge within the standard Blandford-Znajek
scenario, generating electromagnetic emission along dual or single
jets that could be observable to large distance
\cite{Palenzuela:2009yr,Palenzuela:2009hx,Palenzuela:2010nf,Palenzuela:2010xn}
(but see also \cite{Moesta:2011bn,Alic:2012df}).
%\cite{Palenzuela:2011es,Neilsen:2010ax,Palenzuela:2010xn,
%  Palenzuela:2010nf,Moesta:2009rr,Palenzuela:2009hx,Anderson:2009fa,
%  Megevand:2009yx,Palenzuela:2009yr} on Blandford-Znajek. 

%Finally there is
%some work by Zanotti and Rezzolla (e.g.~\cite{Zanotti:2010xs}), but I'm not
%sure it's worth mentioning. Am I missing something?}

%\noindent
%\vc{I think this should not be done:}
%\eb{I squeezed in two citations in a sentence somewhere else, where Uli
%mentioned these ideas in passing.}
%{\bf \em Compact binary census:} Ideas for Earth-based
%\cite{Mandel:2009pc,Mandel:2010xq} and space-based detectors
%\cite{Gair:2010bx,Sesana:2010wy}.

For further reading, we recommend the following reviews.  An excellent summary
of BH mass and spin measurements and (possible) evidence for event horizons
based on ``traditional'' electromagnetic astronomy is given by Narayan
\cite{Narayan:2005ie}.  The use of electromagnetic observations of BHs and
neutron stars to probe strong-field gravity is reviewed by Psaltis
\cite{Psaltis:2008bb}.  A more extended summary of numerical results on
gravitational recoil can be found in Zlochower {\em et al.}
\cite{Zlochower2010}. For a more general review of numerical BH simulations we
recommend Centrella {\em et al.} \cite{Centrella2010,Centrella2010a}.

%=============================================================================
\section{Black holes and high energy physics}
\label{sec:highenergy}
%=============================================================================
The breakthroughs in numerical relativity have opened the door to tackle many
fundamental problems in BH physics and to address questions of wider
interest. A new Golden Era in BH physics is just starting. We
summarize below what we think have been the main developments in this
relatively short period of time.

\vspace{3pt}
\noindent{\bf \emph{Mathematical physics and fundamental issues.}} A
decades-old problem in GR concerns the high-energy collision of two BHs. Due
to the dominance of the gravitational interaction at large energies, this
process is thought to describe general trans-planckian scattering of
particles: at very large center of mass (CM) energies all interactions are
``frozen'' while gravity is boosted, thus all the details about internal
structure of the colliding particles should be washed out. This is further
supported by Thorne's hoop conjecture (recently tested in highly dynamical
situations \cite{Choptuik2009}), which predicts BH formation from generic
high-energy collisions of particles \cite{Thorne:1972ji}. Because the final BH
horizon cloaks all details about the structure of the colliding objects, {\it
  matter does not matter} and one can for simplicity study BHs as representing
a wide class of colliding objects \cite{Giddings2002,Dimopoulos2001}.

For large CM energies, BH collisions are arguably the most violent and
nonlinear process one could conceive of. Evolving Einstein's equations in such
extreme regimes poses new problems, such as the need to deal with all the
different scales involved, and raises several questions. A fundamental issue
concerns Cosmic Censorship. In four spacetime dimensions,
BHs have an upper bound on their angular momentum,
given by\footnote{We note that the Earth's spin $\chi_{\rm Earth}\sim 10^3$,
  while that of a spinning top $\chi_{\rm top}\sim 10^{18}$, so (even though the
  event horizon generators move at the speed of light) BHs rotate ``slowly'',
  as measured by their Kerr parameters.}
\begin{equation}
\chi \equiv Sc/(GM^2)\leq1\,. \label{kerrbound}
\end{equation}
Is it possible to tune the impact parameter in such a way as to produce a
final object spinning above the Kerr bound, or does nature somehow conspire to
always radiate enough angular momentum as to produce a BH?  This problem has
been addressed in recent years
\cite{Sperhake2008,Sperhake2009,Shibata2008,Sperhake2010}.
\begin{figure}[th]
\centering
\includegraphics[width=4.0cm,clip=true]{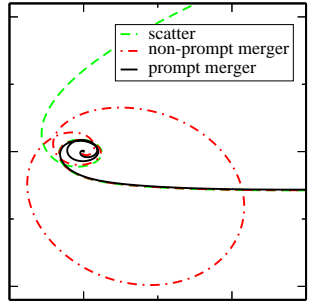}
\caption{Puncture trajectories of one BH for a scattering orbit
  ($b[=3.40M]>b_{\rm scat}$), a prompt merger ($b[=3.34M]<b^*$) and a
  nonprompt merger ($b^*<b[=3.39M]<b_{\rm scat}$). Taken from
  Ref.~\cite{Sperhake2009}.
  %\eb{We should define the thresholds in the main text.}
  }
\label{fig:crit_traj}
\end{figure}

\begin{figure}
\centering
\includegraphics[width=10cm,bb = 12 592 510 782,clip]{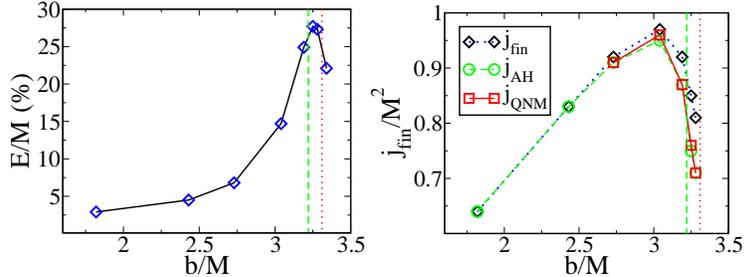}
\caption{Total energy radiated (left) and final BH spin (right) vs. impact
  parameter from sequence 2, the latter calculated using several methods.  The
  vertical dashed green (dotted red) line is the estimated threshold of
  immediate merger $b^*$ (the scattering threshold $b_{\rm scat}$).
}\label{fig:E_b}
\end{figure}

We show the trajectory of two equal-mass, nonspinning,
colliding BHs in Fig.~\ref{fig:crit_traj}, for CM velocities of around
$0.75c$. We notice three distinct regimes.
For impact parameters above the {\em scattering threshold} ($b>b_{\rm scat}$,
the gravitational interaction between the two colliding BHs is relatively
weak and they scatter to
infinity. For impact parameters below the {\em threshold of
immediate merger} ($b<b_*$), the BHs promptly collide and merge into a single BH.
The transition between these two states for $b_*<b<b_{\rm scat}$ is a
nonprompt merger, which puts the BHs in a so-called zoom-whirl orbit,
i.e.~the number of orbits $n_{\rm orb} \sim \ln|b-b_*|$.  The
results show that delayed mergers are necessary in order for the system to
radiate excess angular momentum, so that the end product obeys
(\ref{kerrbound}). The radiated energy and angular momentum increase
strongly as the impact parameter approaches the threshold of immediate merger,
as can be seen in Fig.~\ref{fig:E_b}.
%\vc{should we mention the orbital hang-up effect in this connection?}
%
For {\it head-on} collisions, numerical results suggest that a fraction $14\%$
of the CM energy is radiated in the limit of very large CM energy
\cite{Sperhake2008}. For finely tuned impact parameters, the
total radiated energy (in GWs) can be of order $30\%$ for a collision with
$v\sim 0.75 c$ in the CM frame, while the spin of the final hole can reach
$\simeq 96\%$ of the bound (\ref{kerrbound}). An open issue is how much CM
energy can be radiated for larger boosts.  Finally, peak luminosities of
around $0.1 c^5/G$ are attained in these simulations. This number is orders of
magnitude above any electromagnetic
phenomena and close to the maximum conjectured possible
value \cite{Thorne1983,Gibbons:2002iv}.  The scattering threshold as a
function of boost was studied by Shibata {\it et al.} and Sperhake {\it et
  al.} \cite{Sperhake2009,Shibata2008,Sperhake2010}. For the range of boosts
studied it is well approximated by
\begin{equation}
b_{\rm scat}/M\sim (2.5\pm 0.05)/v\,.
\end{equation}
Here $M$ is the total ADM mass and $v$ the velocity of each hole as measured
in the CM frame.

%This kind of setting is crucial for an understanding of Cosmic Censorhip, but
%also to understand different formulations of Einstein's gravity in four- and
%higher-dimensional spacetimes and their numerical well-posedness properties.
%
%\vc{Refs for this and below missing. Uli, can we add your paper with David and
%  a couple more as the sole refs?}.
%\us{As much as I would like to cite our
%paper with David, I would probably prefer dropping this paragraph. It looks
%a bit of a long shot to add here formulation issues.}
%
%It is also interesting to generalize numerical relativity to alternative
%theories of gravity, which will eventually allow one to test these theories.
%Several problems plague these simulations and hamper the extension to larger
%boosts, the most serious being a strong content of spurious radiation in the
%initial data. Further studies are necessary to resolve this issue.

\vspace{3pt}
\noindent{\bf \emph{Higher-dimensional black holes.}} Although BH solutions
and their properties should in principle fall under the ``mathematical
physics'' category, the activity in this specific field has been so intense
that it deserves a subsection of its own.  Asymptotically flat higher-dimensional black objects have a much richer structure than their four-dimensional counterparts. For instance, spherical topology is not the only
allowed topology for objects with a horizon. One can also have, \emph{e.g.},
black rings, with a donut-like topology. Remarkably, these two different
horizon topologies coexist for certain regions in phase-space
\cite{Emparan2008}.  Explorations of the stability of general
higher-dimensional BHs are in their infancy.  Generically it has been
conjectured that, for $D\ge 6$, ultra-spinning Myers-Perry BHs will be unstable
\cite{Emparan:2003sy}. This instability has been confirmed by an analysis of
linearized axisymmetric perturbations in $D=7,8,9$
\cite{Dias:2009iu}. Clearly, the study of the nonlinear development of these
instabilities requires numerical methods, such as the ones reviewed here. A
study of this type was very recently presented for nonaxisymmetric
perturbations in $D=5$ \cite{Shibata2009,Shibata2010}, where it was found that
a single spinning five dimensional Myers-Perry BH is unstable, for
sufficiently large rotation parameter (confirming previous conjectures
\cite{Cardoso:2006sj,Cardoso:2009bv,Cardoso:2009nz}).

General equilibrium states in anti-de Sitter backgrounds only recently started
to be explored: nonaxisymmetric solutions have finally been built
\cite{Dias:2011at}, confirming previous conjectures about their existence \cite{Cardoso:2009bv,Cardoso:2009nz},
and braneworld BHs are now also starting to be explored
\cite{Figueras:2011gd}. Fully dynamical situations are still uncharted
territory.

\vspace{3pt}
\noindent {\bf \emph{TeV-scale gravity scenarios.}} The above studies are of
direct relevance to some high-energy scenarios.  An outstanding problem in
high-energy physics is the extremely large ratio between the four dimensional
Planck scale, $10^{19}$ GeV, and the electroweak scale, $10^2$ GeV. It has
been proposed that this \textit{hierarchy problem} can be resolved by
confining the Standard Model to a brane in a higher dimensional space
\cite{Arkani-Hamed1998,Antoniadis1998,Randall1999,Randall1999a}.

In such models, the fundamental Planck scale -- the energy at which
gravitational interactions become strong -- could be as low as 1 TeV. Thus,
high-energy colliders, such as the Large Hadron Collider (LHC), may directly
probe strongly coupled gravitational physics
\cite{Argyres:1998qn,Banks1999,Giddings2002,Dimopoulos2001}.  In fact, such
tests may even be routinely available in the collisions of ultra high-energy
cosmic rays with the Earth's atmosphere
\cite{Feng2002,Ahn:2003qn,Cardoso:2004zi}, or in astrophysical BH environments
\cite{Banados:2009pr,Berti2009a,Jacobson:2009zg} (for reviews see
\cite{Cavaglia:2002si,Kanti2004,Kanti:2008eq}).  The production of BHs at
trans-Planckian collision energies (compared to the fundamental Planck scale)
should be well described by using classical GR extended to $D$~dimensions
\cite{Banks1999,Giddings2002,Dimopoulos2001,Cavaglia:2002si,Kanti2004,Kanti:2008eq}. The
challenge is then to use the classical framework to determine the cross
section for production and, for each initial setup, the fractions of the
collision energy and angular momentum that are lost in the higher-dimensional
space by GW emission. This information will be of paramount importance to
improve the modelling of microscopic BH production in event generators.

The first models for BH production in parton-parton collisions used a simple
black disk approach to estimate the cross section for production
\cite{Giddings2002,Dimopoulos2001}. As we already described, only recently
exact results for highly relativistic collisions where obtained in four
dimensions, using numerical relativity techniques
\cite{Sperhake2008,Sperhake2009,Shibata2008,Sperhake2010}.  The extension to
higher-dimensional spacetimes is a topic of current investigations
\cite{Yoshino2009,Zilhao2010,Witek2010a}. The first numerical results
concerning low-energy collisions have been reported last year
\cite{Witek2010a,Witek2010c}, initial data for boosted
BH binaries have been constructed using an extended puncture
approach in \cite{Yoshino:2006kc,Zilhao:2011yc}
and some results for high-energy collisions have
been presented in \cite{Okawa:2011fv}.

\vspace{3pt}
\noindent{\bf \emph{AdS/CFT and holography.}}  
In 1997--98, a powerful new technique known as the AdS/CFT correspondence or,
more generally, the gauge-gravity duality, was introduced and rapidly
developed \cite{Maldacena1997}. This holographic correspondence provides an
effective description of a nonperturbative, strongly coupled regime of
certain gauge theories in terms of higher-dimensional classical gravity in AdS backgrounds. In
particular, equilibrium and nonequilibrium properties of strongly coupled
thermal gauge theories are related to the physics of higher-dimensional BHs,
black branes and their fluctuations. These studies revealed intriguing
connections between the dynamics of BH horizons and hydrodynamics, and offer
new perspectives on notoriously difficult problems, such as the BH information
loss paradox, the nature of BH singularities and quantum gravity.

Numerical relativity in anti-de Sitter backgrounds is bound to contribute
enormously to our understanding of the gauge-gravity duality and is likely to
have important applications in the interpretation of observations
\cite{Mateos:2007ay,Hartnoll:2009sz,Amsel2007,Gubser2008}.  For instance, in
the context of the gauge-gravity duality, high-energy collisions of BHs have a
dual description in terms of \textit{a)}~high-energy collisions with balls of
deconfined plasma surrounded by a confining phase, and \textit{b)}~the rapid
localized heating of a deconfined plasma. These are the type of events that
may have direct observational consequences for the experiments at Brookhaven's
Relativistic Heavy Ion Collider (RHIC) \cite{Amsel2007,Gubser2008}. Numerical
relativity in anti-de Sitter is particularly difficult, and so far only very
special situations have been handled
\cite{Pretorius:2000yu,Witek2010,Chesler:2008hg,Chesler:2010bi,
Bantilan:2012vu}.
Among these, we note the recent numerical proof that pure AdS space is unstable against collapse to BHs
\cite{Bizon:2011gg,Jalmuzna:2011qw}.

BHs in the context of high-energy physics have also been discussed by
Pretorius \cite{Pretorius2007a} and in the white paper \cite{Cardoso:2012qm}.
A recent review of numerical simulations of black
strings in five spacetime dimensions is given by Lehner and Pretorius
\cite{Lehner2011}. An overview about numerical results on
BHs in $D>4$ dimensions
is given in \cite{Yoshino:2011zz,Yoshino:2011zza}.
Finally we note Kanti's review on BHs at the LHC
\cite{Kanti:2008eq}.

%%%%%%%%%%%%%%%%%%%%%%%%%%%%%%%%%%%%%%%%%%%%%%%%%%%%%%%%%%%%%%%%%%%%%%%%%%%%%%
\section{Conclusions}
\label{sec:conclusions}
%In summary, numerical relativity has generated a wealth of results on the
%dynamics of BHs, in particular BH binary systems. The
%construction of GW template banks for use in GW data analysis is
%under way with increasingly accurate numerical waveforms that are being
%used to calibrate phenomenological and EOB waveform models. Numerical
%studies have revealed partly surprising results about astrophysical
%systems involving BHs, most notably the large magnitude of
%the so-called superkicks. Analytic approximations for the kicks and
%final spins resulting from BH merger have been constructed and
%will be improved in the future. They play a vital role for understanding
%the formation history of massive BHs and what type of
%BH populations we should expect to encounter in our searches
%of the Universe. Most recently, the area of numerical relativity has
%been extended to connect to high-energy physics; most importantly
%to improve our modeling of trans-planckian scattering of particles,
%but also to simulate BHs in more generic classes of spacetimes
%and thus explore BH dynamics in the context of the AdS/CFT
%correspondence.

We conclude this review with a brief summary of the most urgent problems to be
tackled by numerical relativity in the three main areas in the near future.

\vspace{3pt}
\noindent
{\bf \em Gravitational wave physics:} In order to meet tight accuracy
requirements in GW template generation, especially for parameter estimation,
we either need longer numerical waveforms or a denser spacing of templates in
the parameter space. Both approaches will require computational resources that
grow non-linearly as a function of the accuracy, either because of the slow
nature of the inspiral at larger BH separations (see Sec.~2 in
\cite{MacDonald2011} for a quantitative discussion) or because of the dimensionality of
the parameter space.  A second major task is the extension of existing
template models to generic binaries with precessing spins and/or smaller
mass ratios, bridging the gap to perturbative modeling of extreme mass ratio
binaries.

\vspace{3pt}
\noindent
{\bf \em Astrophysics:} While existing formulae for the kick and final spin
resulting from the coalescence of BHs have been helpful for astrophysical
studies, the calibration of generic models as proposed by Boyle {\em et al.}
\cite{Boyle2007a,Boyle2007b} requires a more comprehensive exploration of the
parameter space. The relatively young field of numerical relativity
simulations of BHs surrounded by accretion disks will likely
improve our insight into expected optical counterparts to BH binary
mergers, and thus provide a vital tool for the area of {\em multi-messanger
astrophysics}.

\vspace{3pt}
\noindent
{\bf \em High-energy physics:} Arguably the most urgent task is the
determination of the scattering cross-section and the loss of energy and
momentum in GWs in trans-Planckian scattering of BHs in {\em arbitrary}
dimensions. In view of the stability problems encountered in present studies, it
also appears desirable to obtain a better understanding of the well-posedness
of the formulations currently employed for such studies. Second, the modeling
of BHs in generic spacetimes such as de Sitter and anti-de Sitter is still in
its infancy, even if preliminary studies of BHs in non-asymptotically flat
spacetimes are encouraging. A better understanding of boundary issues,
well-posedness of the formulations and diagnostics such as wave extraction
tools and horizon finders will be required to open up this uncharted and
fertile ground of research.

\vspace{5pt}
\noindent
{\bf Acknowledgements:} U.S. acknowledges support from the Ram{\'o}n y Cajal
Programme and Grant No.~FIS2011-30145-C03-03
of the Ministry of Education and Science of Spain,
the FP7-PEOPLE-2011-CIG Grant {\em CBHEO 293412},
NSF grants
PHY-0601459, PHY-0652995 and the Sherman Fairchild Foundation to Caltech.
E.B.'s research is supported by NSF grant PHY-0900735
and by NSF CAREER grant PHY-1055103.
This work was supported by the DyBHo-256667 ERC Starting Grant,
%by the FP7-PEOPLE-2011-IRSES Grant {\em NRHEP 295189}
and by FCT-Portugal through
projects PTDC/FIS/098025/2008, PTDC/FIS/098032/2008, CTE-AST/098034/2008,
CERN/FP/116341/ 2010 and CERN/FP/123593/2011. This research was supported by
the DyBHo--256667 ERC Starting Grant, the NRHEP--295189
FP7--PEOPLE--2011--IRSES Grant
by allocations through the
TeraGrid Advanced Support Program at the San Diego Supercomputing Center and
the National Institute for Computational Sciences under grant PHY-090003, the
Centro de Supercomputaci{\'o}n de Galicia (CESGA) under project numbers
ICTS-CESGA-175, ICTS-CESGA-200 and ICTS-CESGA-221,
the Barcelona Supercomputing Center (BSC)
under project AECT-2011-2-0006, AECT-2011-3-0007, AECT-2012-1-0008,
the DEISA Extreme Computing Initiatives DECI-6 and
DECI-7 PRACE-2IP Project DyBHO.

%% The Appendices part is started with the command \appendix;
%% appendix sections are then done as normal sections
%% \appendix

%% \section{}
%% \label{}

%% References
%%
%% Following citation commands can be used in the body text:
%% Usage of \cite is as follows:
%%   \cite{key}         ==>>  [#]
%%   \cite[chap. 2]{key} ==>> [#, chap. 2]
%%

%% References with bibTeX database:

\bibliographystyle{elsarticle-num}
%\bibliography{uli_short}

%% Authors are advised to submit their bibtex database files. They are
%% requested to list a bibtex style file in the manuscript if they do
%% not want to use elsarticle-num.bst.

%% References without bibTeX database:

% \begin{thebibliography}{00}

%% \bibitem must have the following form:
%%   \bibitem{key}...
%%

% \bibitem{}

% \end{thebibliography}

\end{document}